\documentclass{article}
\usepackage[utf8]{inputenc}
\usepackage{amsmath}
\usepackage{amssymb}
\usepackage{booktabs}
\usepackage{graphicx}
\usepackage{caption}
\usepackage{subcaption}
\usepackage{adjustbox}
\usepackage[capitalise]{cleveref}
\usepackage{geometry}
\usepackage[normalem]{ulem}
\useunder{\uline}{\ul}{}
\usepackage{multirow}

\usepackage{booktabs}
\usepackage[table,xcdraw]{xcolor}

\usepackage[table,xcdraw]{xcolor}

\title{Bounded strategic reasoning explains crisis emergence in multi-agent market games
}
\author{
Benjamin Patrick Evans\footnote{benjamin.evans@sydney.edu.au} \and
Mikhail Prokopenko
}

\date{\footnotesize Centre for Complex Systems, The University of Sydney, Sydney, NSW {2006}, Australia}

\begin{document}

\maketitle

\begin{abstract}
The efficient market hypothesis (EMH), based on rational expectations and market equilibrium, is the dominant perspective for modelling economic markets. However, the most notable critique of the EMH is the inability to model periods of out-of-equilibrium behaviour in the absence of any significant external news. When such dynamics emerge endogenously, the traditional economic frameworks provide no explanation for such behaviour and the deviation from equilibrium. This work offers an alternate perspective explaining the endogenous emergence of punctuated out-of-equilibrium dynamics based on bounded rational agents. In a concise market entrance game, we show how boundedly rational strategic reasoning can lead to endogenously emerging crises, exhibiting fat tails in ``returns". We also show how other common stylised facts of economic markets, such as clustered volatility, can be explained due to agent diversity (or lack thereof) and the varying learning updates across the agents. This work explains various stylised facts and crisis emergence in economic markets, in the absence of any external news, based purely on agent interactions and bounded rational reasoning.
\end{abstract}

\section{Introduction}
Economic markets have existed for millennia, with the earliest identified markets dating back to at least the Babylonian Empire \cite{casson2011origin}. Since then, the modelling of such markets has been a heavily researched topic, aiming to improve understanding of markets, increase profits, and shape policy and interventions within these markets. The dominant perspective which has arisen over the last two centuries \cite{sewell2011history} is the efficient market hypothesis (EMH) \cite{fama1970efficient}, based on rational expectations and market equilibrium. The EMH states that prices reflect all known information about an asset, and as such, they reflect the fair value. However, there are multiple significant market crashes, such as the 1987 Dow Jones index crash \cite{danielsson2003endogenous}, the global financial crisis in 2008 \cite{ball2009global}, and the flash crash of 2010 \cite{kirilenko2017flash}, that the EMH failed to explain purely by the arrival of external news. If prices reflected all information, assets matched their actual underlying value, and all agents were perfectly rational and not speculative, such bubbles (and resulting bursts) would not occur. Instead, agents often display irrational exuberance \cite{exuberance}, based on unfounded beliefs and speculation \cite{shiller2000irrational}. 

The most serious specific critiques of the EMH include its inability to explain periods of out-of-equilibrium behaviour such as endogenous crises \cite{buchanan2009meltdown, farmer2009economy} or the ``stylised facts" present in economic markets such as volatility clustering where periods of high (low) volatility occur in bursts, in the absence of any significant external news \cite{hommesStyilized}. Such crises are known to occur in different markets, from housing markets \cite{evans2021maximum, evans2021impact}, to stock markets \cite{song20222020}, and foreign exchange markets \cite{wehrli2022classification}. Likewise, volatility clustering is consistently observed in markets  \cite{mandelbrot1997variation}, with temporal correlations in volatility breaching the traditional economic assumption of heteroskedasticity. The EMH suggests that such phenomena occur due to rational agents reacting to external news in the market \cite{inoua2020news}. However, when such dynamics emerge endogenously, i.e., in the absence of external news, the traditional economic frameworks struggle to explain such behaviour or the deviation from equilibrium \cite{buchanan2017economists}, demanding adequate models explaining the ``wildness" of market dynamics~\cite{bouchaud2008economics, arthurComplexity,harre2021complexity}.

Thus, the endogenous emergence of crises and stylised facts are significant aspects to consider in the modelling of economic markets. An adequate model of economic markets should capture and ideally explain three desirable aspects of actual market dynamics: (i) \textit{convergence} to a good average outcome, i.e., equilibrium, (ii) endogenous \textit{emergence of crises} so that the resulting dynamics are categorised by fat-tailed distributions of out-of-equilibria deviations, where abrupt periods differ significantly from the ``converged" equilibrium, and (iii) ability to recreate \textit{stylised facts} of economic markets such as clustered volatility. The EMH explains (ii) and (iii) as arising only from the arrival of external news.

These notable deviations from EMH have pushed for a ``revolution" of economics \cite{bouchaud2008economics}, and encouraged a modelling focus on agents who operate under ``bounded rationality" \cite{simon1956rational}. Relaxing the perfect rationality assumption provides an alternate perspective to markets, admitting that market participants may not follow the \textit{homo economicus} model \cite{levitt2008homo, camerer2006does}, 
and may instead be speculative, adaptive, or subject to limitations in their information processing abilities. Relaxing this assumption allows for behavioural realism, for example, with heterogeneous agents for modelling the wide range of human reasoning abilities. 
These considerations have given rise to \textit{Complexity Economics} \cite{arthur2021foundations}, and the ``interacting agent hypothesis" \cite{kaizoji2000speculative} capturing the endogenous emergence of out-of-equilibrium dynamics due to the interaction among heterogeneous boundedly rational agents \cite{hommes2002modeling, alfarano2005estimation, schmitt2017herding}, rather than purely by the arrival of external news.

One of the most notable examples exploring Complexity Economics is the canonical El Farol bar problem \cite{arthur1994inductive}. El Farol has been called ``the most important problem" in the modelling of complex systems \cite{casti1996seeing}, and continues to be explored \cite{sellers2020simulating, st2021network}, having motivated a host of other market entrance games \cite{camerer2004cognitive} and minority games \cite{challet2001stylized} for modelling  markets. In this market entrance game the agent payoffs depend on other market participants' decisions, creating complex market dynamics. Adaptive strategies is the widely accepted solution to the El Farol bar problem \cite{arthur1994inductive}, which generally converges to an equilibrium near the optimal resource capacity \cite{challet2004shedding}. However, it is unknown to what extent deviations from this equilibrium can be captured or whether current Complexity Economics solutions to El Farol can generate stylised facts similar to those observed in actual markets. Hence, we need a more refined model capable of demonstrating convergence to equilibrium, punctuated by abrupt deviations, while also explaining market dynamics and stylised facts arising endogenously rather than relying on the arrival of external news. 

Here, we propose an approach based on boundedly rational strategic (higher-order) reasoning agents.
Each Bounded Rational AdapTive Strategic (BRATS) reasoning agent maintains a recursive model about other agents' beliefs. However, the potentially infinite chain of strategic reasoning is ``broken'' at various points of recursion, following the Quantal Hierarchy (QH) model~\cite{evans2021bounded}. Specifically,
the heterogeneous agents are limited in the amount of information processing they can perform, with reasoning resources quantified information-theoretically (in the Shannon sense). 
The BRATS agents learn and update their recursive beliefs (increase resources) based on the observed market outcomes. The BRATS approach improves upon the canonical solution driven by adaptive strategies across all outlined criteria. 
Specifically, the approach not only ensures (i) general convergence towards an ``equilibrium" even with boundedly rational agents, but also allows for (ii) the explanation of abrupt endogenously emerging periods of out-of-equilibrium behaviour based on boundedly rational strategic reasoning, generating ``fat tails". Furthermore, (iii) volatility clustering can be generated and explained endogenously due to the diversity of agent beliefs and the heterogeneous learning updates.

We provide comparisons to alternative approaches and show that the proposed approach convincingly outperforms the alternatives across the key measures in a market game, better matching  dynamics observed in actual economic markets. Thus, the long-held conjecture that the adaptive strategies approach provides an adequate resolution to the El Farol problem is challenged. Surprisingly, a convincing solution is shown to be provided by the agents which reason strategically (i.e., recursively) while being limited by their information-processing resources.

\section{Results}\label{secGame}
The basic premise of the El Farol market entrance game is as follows. Agents enjoy attending a bar, or profit entering a market, if less than $60\%$ of other agents attend the bar, otherwise, the bar/market is deemed overcrowded/unprofitable, and they would have preferred to stay out. As a generalization, we say $N$ agents receive payoff $U_{enter}$ for entering if less than $cN, c \in [0,1]$ other agents attend, otherwise they receive payoff $U_{overcrowded}$. Staying out returns a fixed payoff of  $U_{exit}$. The inequality $U_{enter} > U_{exit} > U_{overcrowded}$ holds such that if the bar is overcrowded, the agent would have preferred to stay home. However, the preference is always to attend if it is not overcrowded. 

\subsection{Convergence}

\begin{figure}[!htb]
    \centering
    \begin{subfigure}[b]{0.45\textwidth}
    \includegraphics[width=\textwidth]{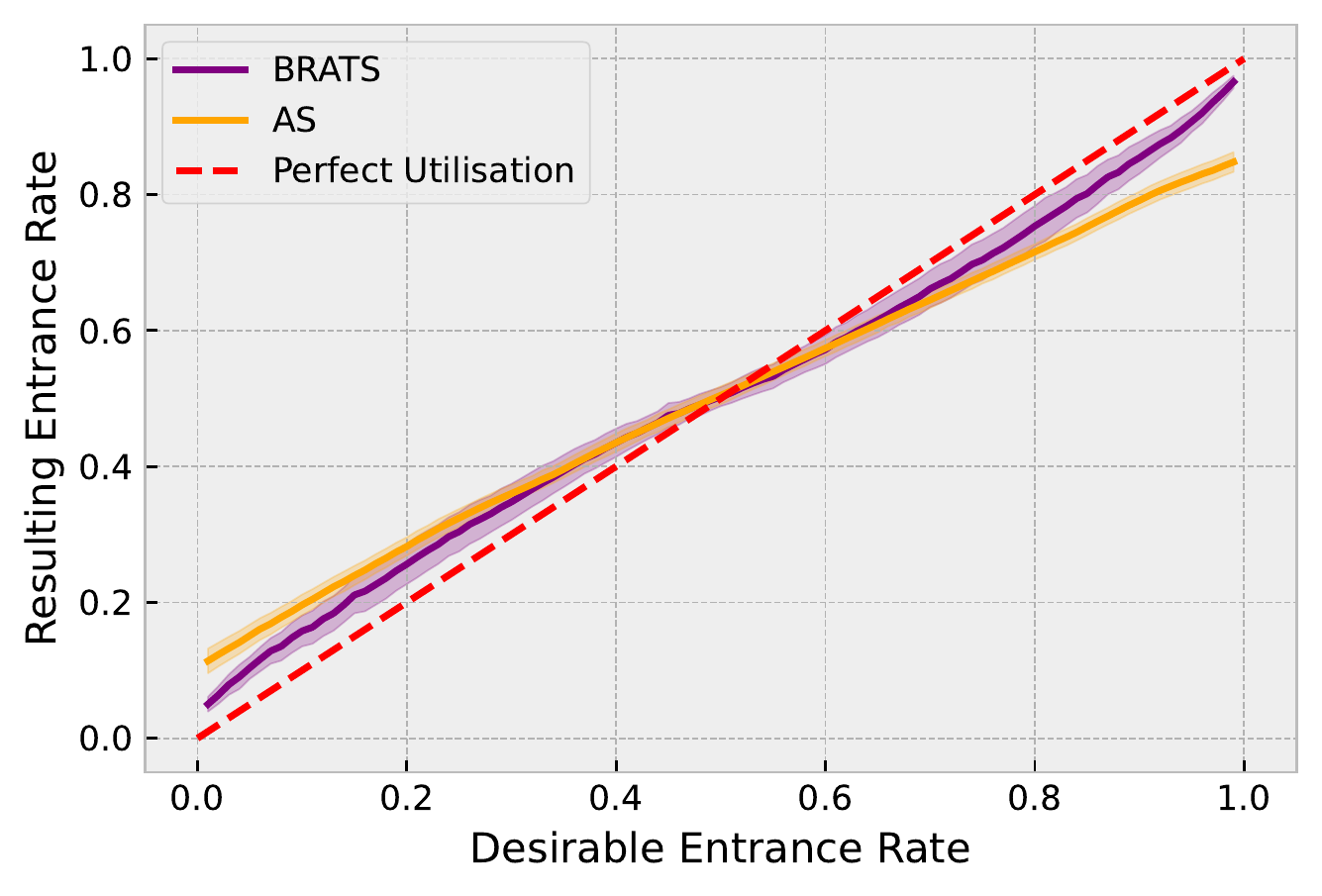}
    \caption{Utilisation}\label{figAttendanceRatesPlot}
    \end{subfigure}
    \begin{subfigure}[b]{0.45\textwidth}
    \includegraphics[width=\textwidth]{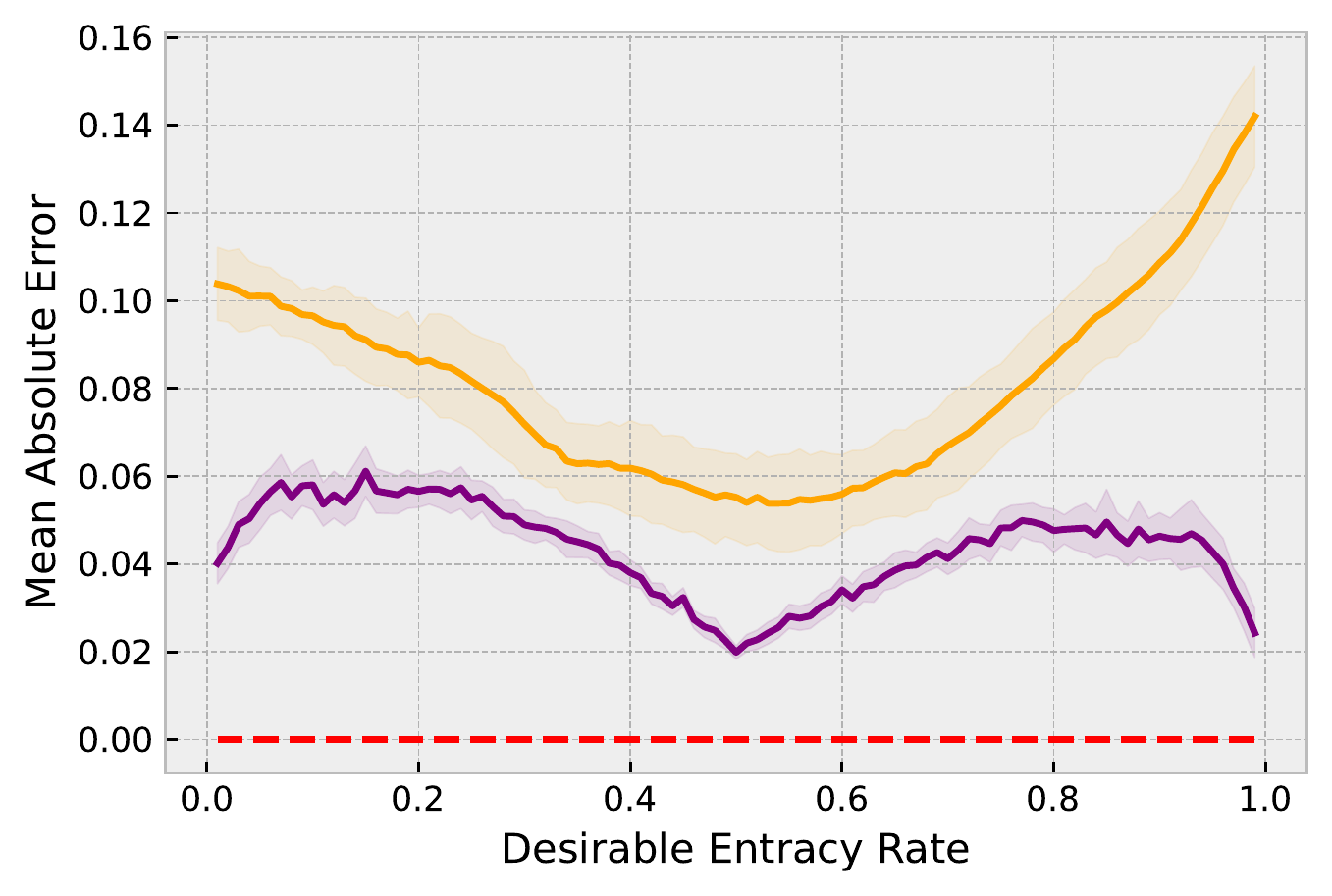}
    \caption{Errors}\label{figAttendanceRatesErrors}
    \end{subfigure}
    \caption{Resource efficiency ($N=100$). The purple (orange) line represents the proposed BRATS model (adaptive strategies). The dotted red line highlights the optimal case, with perfect efficiency, i.e., the mixed strategy Nash equilibrium. \cref{figAttendanceRatesPlot} shows the average overall attendance rate across runs, and the shaded surrounding area shows $\pm$ one standard deviation. \cref{figAttendanceRatesErrors} shows the average error for each run, and the filled area shows $\pm$ one standard deviation across these runs.
    }
    \label{figAttendanceRates}
\end{figure}

We first consider the ``desirable" attendance rate $c$, and the simulated attendance rate from the bounded rational agents. This convergence is displayed in \cref{figAttendanceRatesPlot},  comparing the proposed (BRATS) approach with the canonical adaptive strategies (AS) solution. Agents are able to self-organise around the desired value without explicit coordination. We refer to this as the resource efficiency, i.e., the ability to operate near the optimal capacity. \cref{figAttendanceRatesErrors} shows the resulting errors, with lower average errors for the proposed approach when compared to adaptive strategies.

\begin{figure}[!htb]
     \centering
     \begin{subfigure}[b]{0.3\textwidth}
         \centering
         \includegraphics[width=\textwidth]{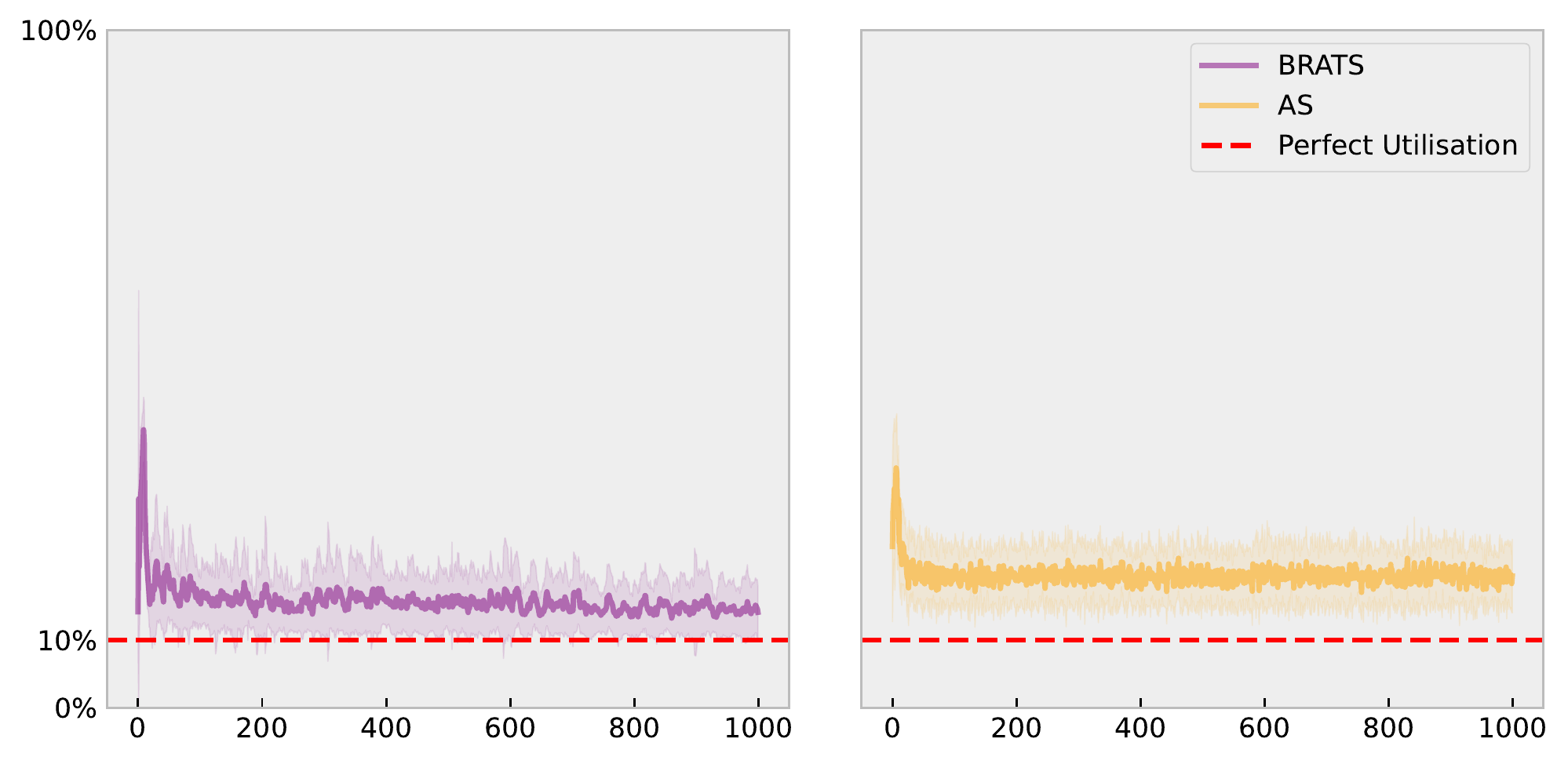}
         \caption{$c=0.1$}
     \end{subfigure}
     \hfill
     \begin{subfigure}[b]{0.3\textwidth}
         \centering
         \includegraphics[width=\textwidth]{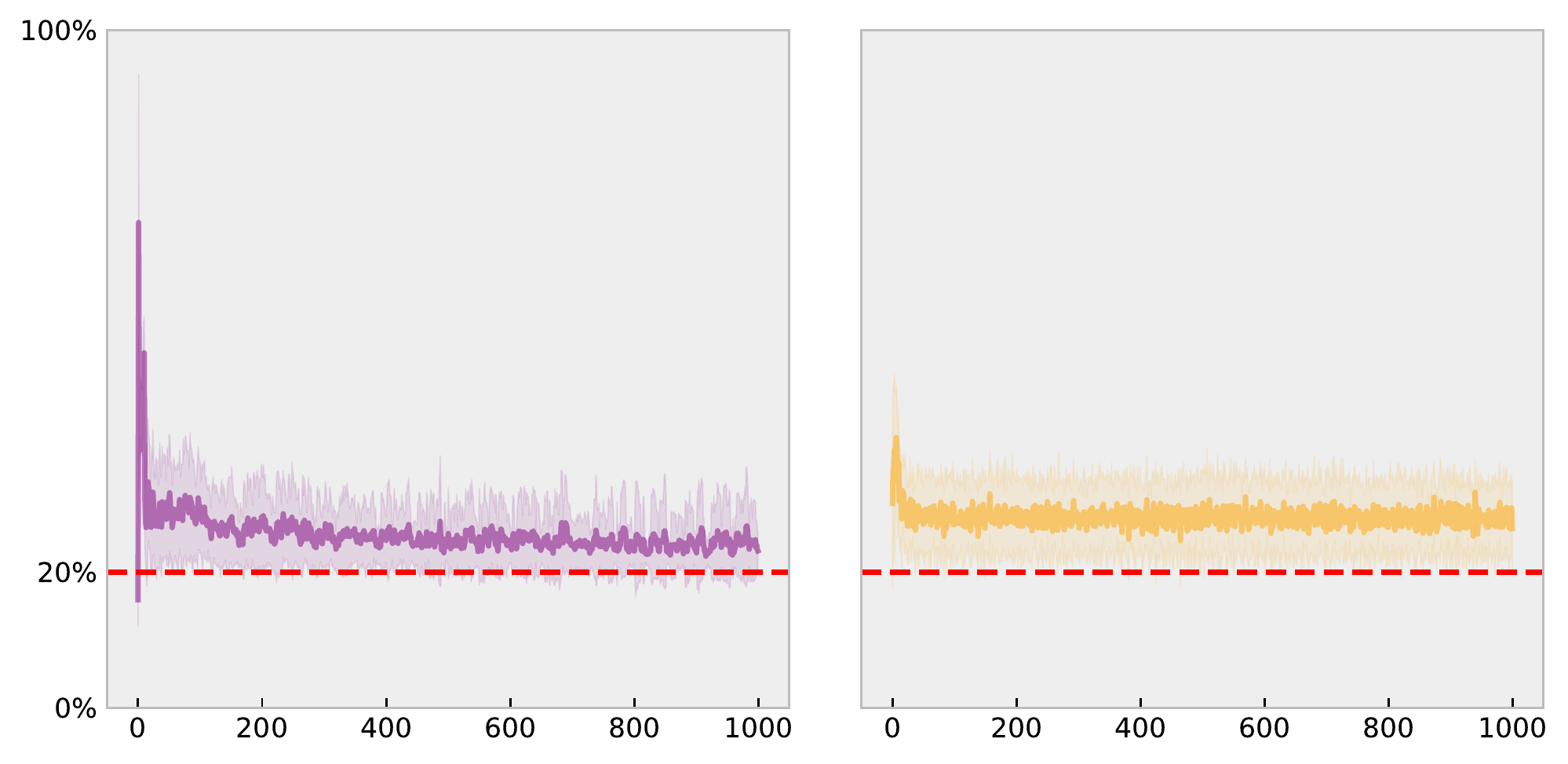}
         \caption{$c=0.2$}
     \end{subfigure}
     \hfill
     \begin{subfigure}[b]{0.3\textwidth}
         \centering
         \includegraphics[width=\textwidth]{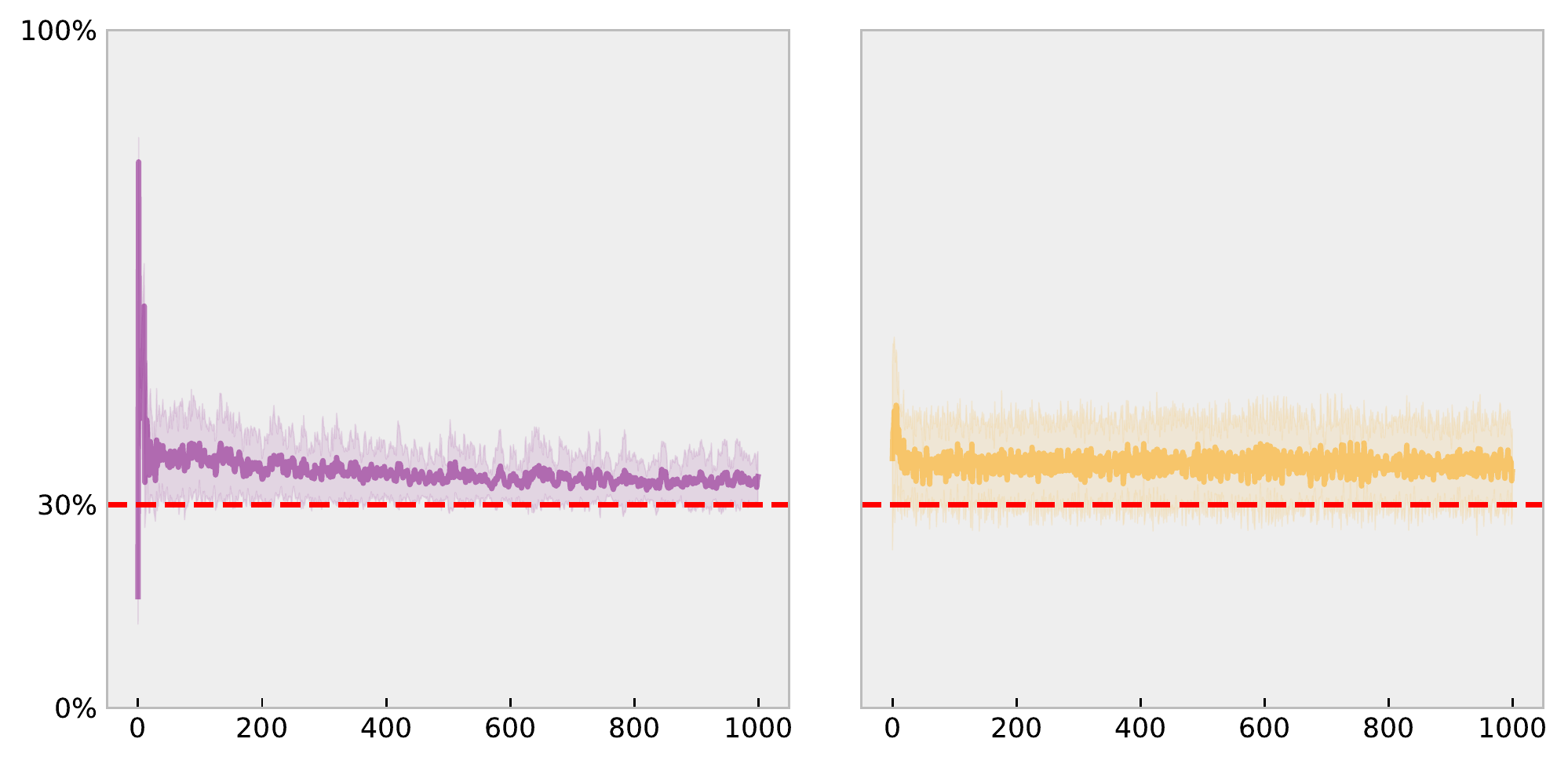}
         \caption{$c=0.3$}
     \end{subfigure}
     \hfill
          \begin{subfigure}[b]{0.3\textwidth}
         \centering
         \includegraphics[width=\textwidth]{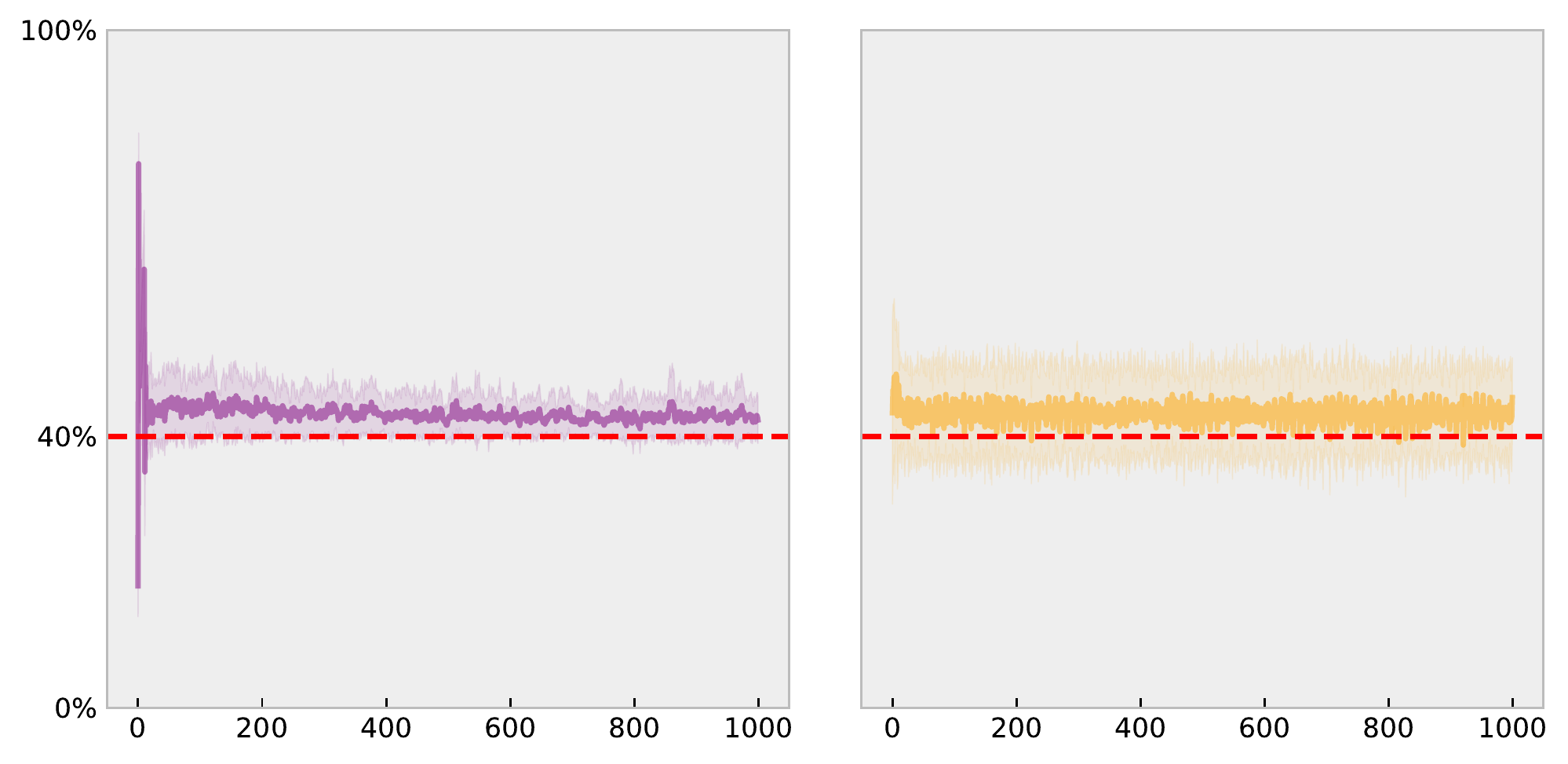}
         \caption{$c=0.4$}
     \end{subfigure}
     \hfill
     \begin{subfigure}[b]{0.3\textwidth}
         \centering
         \includegraphics[width=\textwidth]{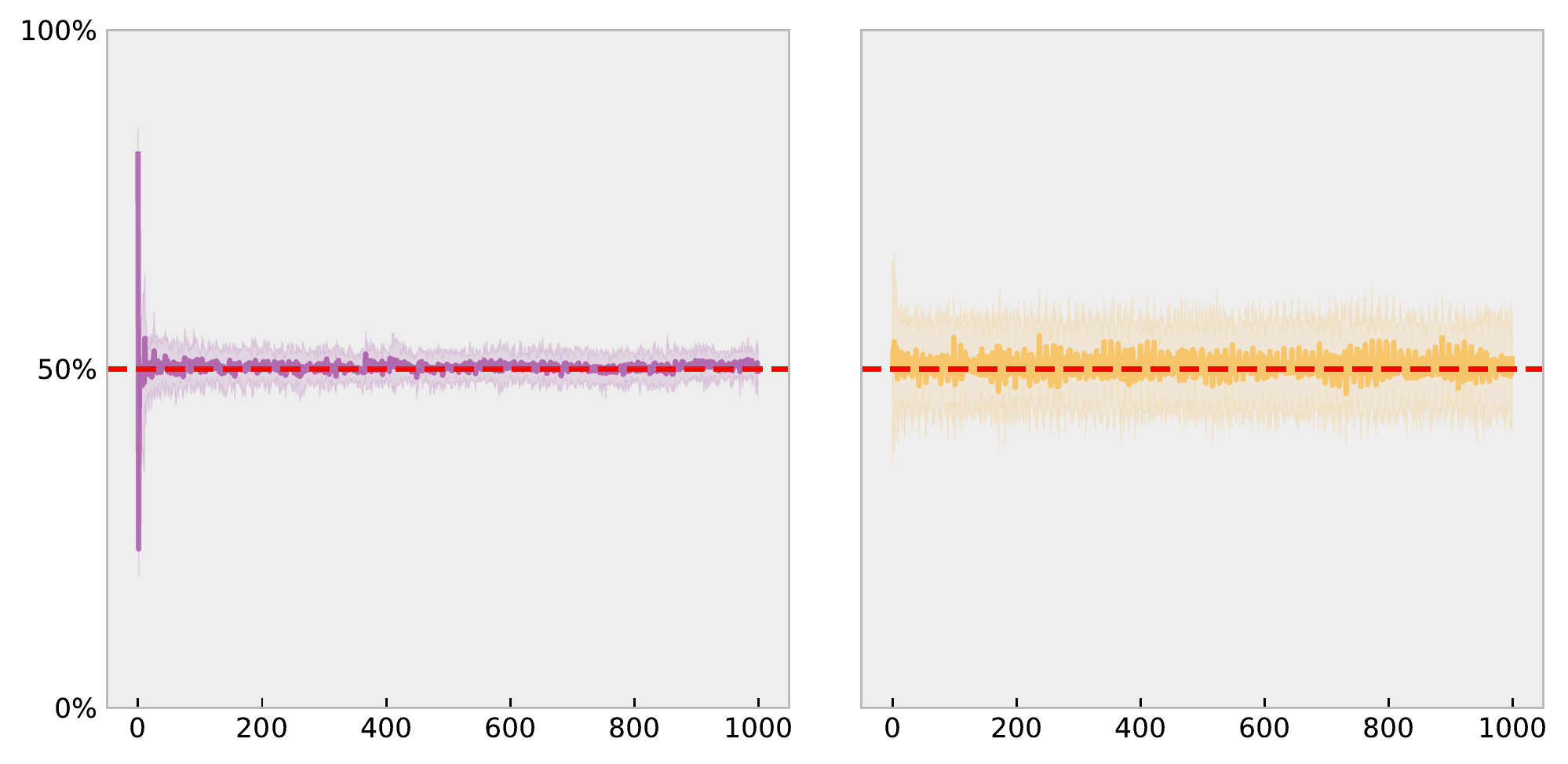}
         \caption{$c=0.5$}
     \end{subfigure}
     \hfill
     \begin{subfigure}[b]{0.3\textwidth}
         \centering
         \includegraphics[width=\textwidth]{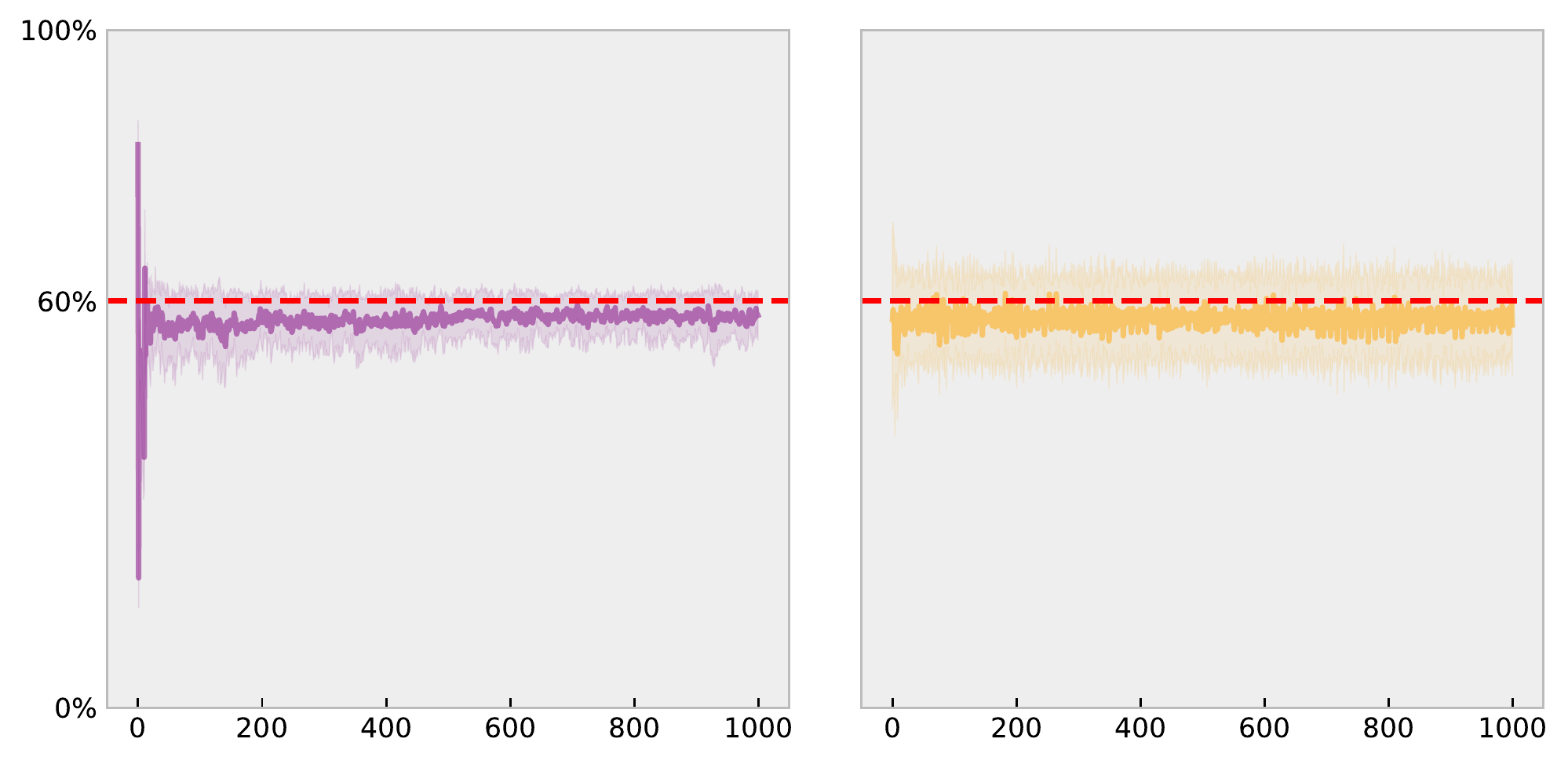}
         \caption{$c=0.6$}
     \end{subfigure}
     \hfill
          \begin{subfigure}[b]{0.3\textwidth}
         \centering
         \includegraphics[width=\textwidth]{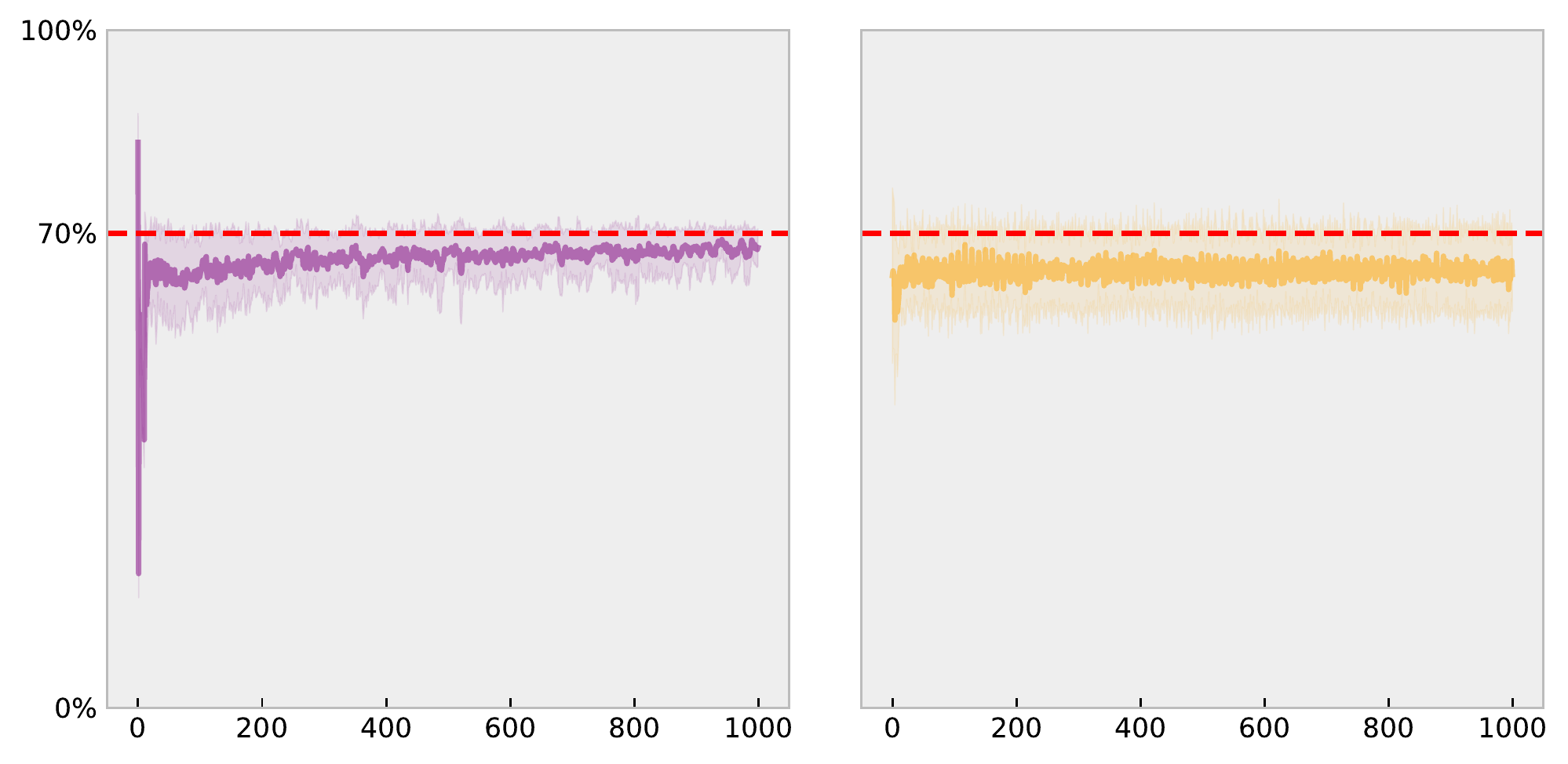}
         \caption{$c=0.7$}
     \end{subfigure}
     \hfill
     \begin{subfigure}[b]{0.3\textwidth}
         \centering
         \includegraphics[width=\textwidth]{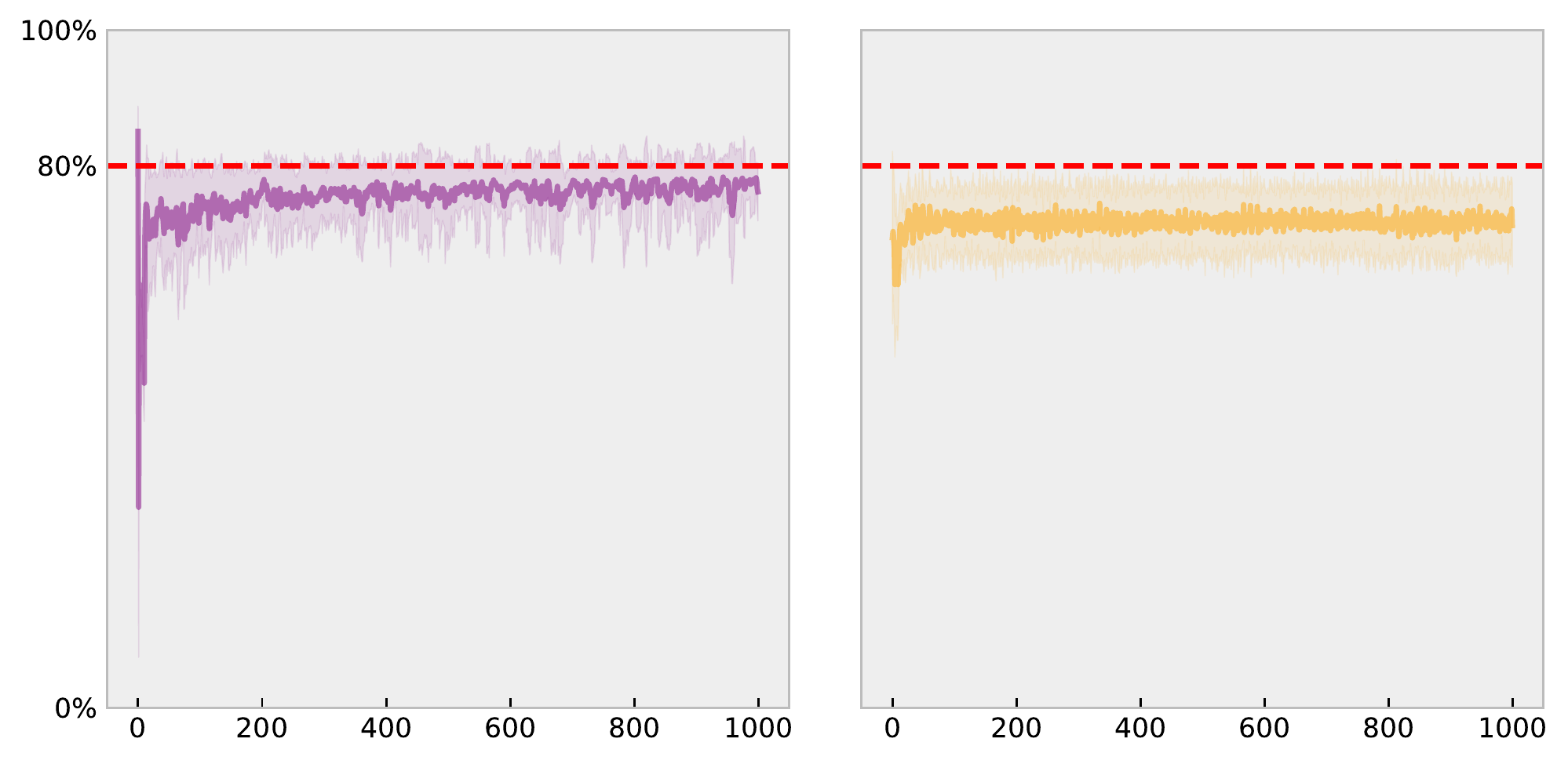}
         \caption{$c=0.8$}
     \end{subfigure}
     \hfill
     \begin{subfigure}[b]{0.3\textwidth}
         \centering
         \includegraphics[width=\textwidth]{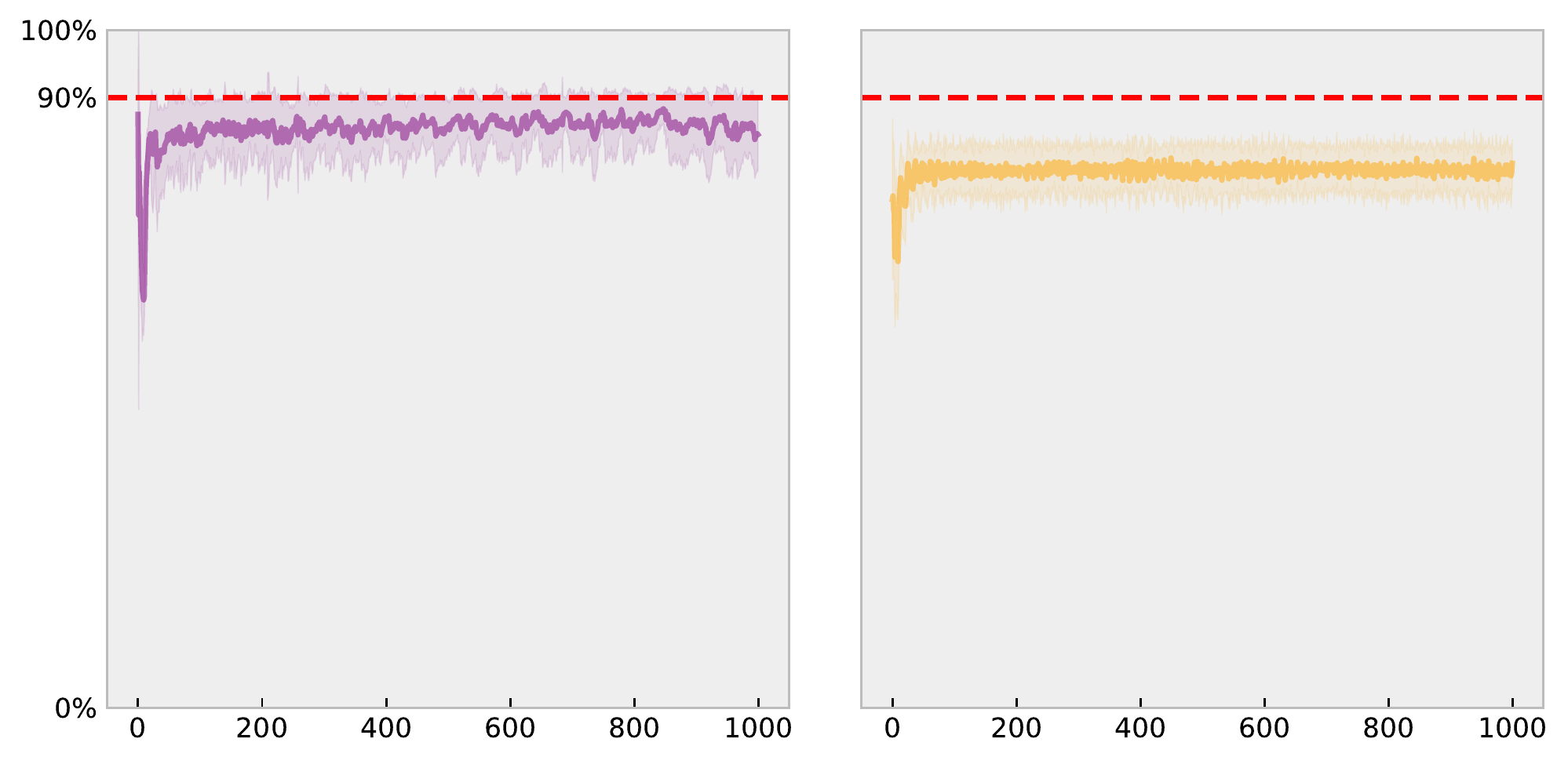}
         \caption{$c=0.9$}
     \end{subfigure}
        \caption{Average resource utilisation for various enjoyable capacities $c$.
        Purple (orange) represents the proposed BRATS model (adaptive strategies). The mean attendance is shown as the dark line, with $\pm$ one standard deviation as the lighter bars. Perfect utilisation ($c$) is displayed as the dashed red line.}
        \label{figSimulationAttendance}
\end{figure}

The time evolution of the system is displayed in \cref{figSimulationAttendance}. There is an initial ``adjustment" period with both approaches in which the agents are learning appropriate beliefs. However, once some learning has taken place (e.g. several rounds have been completed), the proposed BRATS approach converges to a higher average resource efficiency than the canonical AS approach, particularly for higher or lower values of $c$, as confirmed in \cref{figAttendanceRates}. The methods perform similarly for mid ranges of $c$, achieving high resource efficiency. These results show that both approaches perform well on the first criteria of  \textit{convergence} to a good average outcome, with the proposed approach outperforming adaptive strategies for higher and lower entrance capacities. 

It is well known that there exists a unique symmetric mixed-strategy Nash equilibrium (MSE) solution to the problem, where agents attend probabilistically based on the enjoyable capacity $c$. Such an approach offers perfect convergence and utilisation (i.e. $0$ error), however, this MSE solution can not categorise learning or adaption throughout time \cite{shubik2011farol}, and further, can not generate periods of endogenously emerging crises. This limitation is problematic for explaining known deviations from equilibrium in actual market settings. As we are particularly interested in out-of-equilibrium dynamics generated from bounded rational agents, we do not explore such solutions further as they cannot capture such phenomena. For a complete discussion on deriving equilibria in El Farol, see \cite{whitehead2008farol}. In the following sections, these deviations from equilibrium are analysed in more detail.

\subsection{Emergence of endogenous crises}

The previous subsection showed the average convergence towards the desired resource capacity $c$. Here, we consider the endogenous emergence of crises, or self-induced shocks  \cite{kay2004memory}, as a result of endogenous changes in agent beliefs \cite{guzman2020towards}. Following \cite{kay2004memory}, we consider crises to occur when the resulting attendance change is at least three standard deviations $>3\sigma$ from the average historical attendance change, and refer to these as $3^{+}\sigma$ events. The EMH suggests these deviations would follow a random walk, and, thus, be normally distributed, meaning that the occurrence of $3^{+}\sigma$ events would be small ($0.3\%$). However, in actual markets, the kurtosis of the observed return distribution can often be categorised as leptokurtic, implying fatter tails, and a higher probability of ``crisis" than the probability suggested by the random walk, casting doubts about the explanation given by the EMH \cite{jefferies2003anatomy, johnson2003financial}. 

To analyse the frequency of significant changes, we calculate the probability of $3^+\sigma$ events, and in addition, estimate the tail-index $\alpha$ of these changes using the Hill estimator \cite{hill1975simple}. The tail-index $\alpha$ gives a measure of the shape of the tail, with lower $\alpha$'s implying a slower decaying function and thus, heavier tails. It has been reported   that for many markets, $\alpha$ can often be in the range $1$ to $4$~\cite{lebaron2005extreme} with emerging markets having estimates $1<\alpha<2$, and developed markets $2<\alpha<4$~\cite{lebaron2005extreme, lux2000volatility}.

Thus, this section demonstrates endogenous crisis formation in terms of fat tails in attendance fluctuations, supported by estimates of the tail-indexes $\alpha$.


\begin{table}[!htb]
\centering
\begin{tabular}{@{}rccccccccc@{}}
\toprule
\rowcolor[HTML]{F5F5F5} & \multicolumn{9}{c}{c}                      \\
\rowcolor[HTML]{F5F5F5} 
               & \textbf{0.1} & \textbf{0.2} & \textbf{0.3} & \textbf{0.4} & \textbf{0.5} & \textbf{0.6} & \textbf{0.7} & \textbf{0.8} & \textbf{0.9}\\
\midrule
\textbf{BRATS} & 2.3     & 2       & 1.3     & 0.7     & 1       & 1.1     & 1.3     & 1.9     & 2.3     \\
AS             & 1.5     & 1.4     & 1.1     & 1       & 0.8     & 0.6     & 0.5     & 0.4     & 0.6     \\
Noise Traders         & 0.2     & 0.3     & 0.3     & 0.2     & 0.3     & 0.3     & 0.3     & 0.3     & 0.3     \\ \midrule
{\color[HTML]{343434} \textit{Normal Distribution}}  & {\color[HTML]{343434} 0.3}    & {\color[HTML]{343434} 0.3}    & {\color[HTML]{343434} 0.3}      & {\color[HTML]{343434} 0.3}    &{\color[HTML]{343434} 0.3}     & {\color[HTML]{343434} 0.3}      & {\color[HTML]{343434} 0.3}     & {\color[HTML]{343434} 0.3}     & {\color[HTML]{343434} 0.3}      \\ \bottomrule
\end{tabular}
\caption{The occurrence of $3^{+}\sigma$ events. The table displays the observed percentage of attendance changes falling outside 3 standard deviations of the mean attendance. Each cell represents the average results for the attendance rate $c$. According to the normal distribution (bottom row), the percentage of expected occurrences for $3^{+}\sigma$ events is $0.3$, and any values greater than this threshold indicate a higher probability of extreme events.}\label{tblAttendanceRanges}
\end{table}

\begin{figure}[!htb]
    \centering
    \includegraphics[width=.5\textwidth]{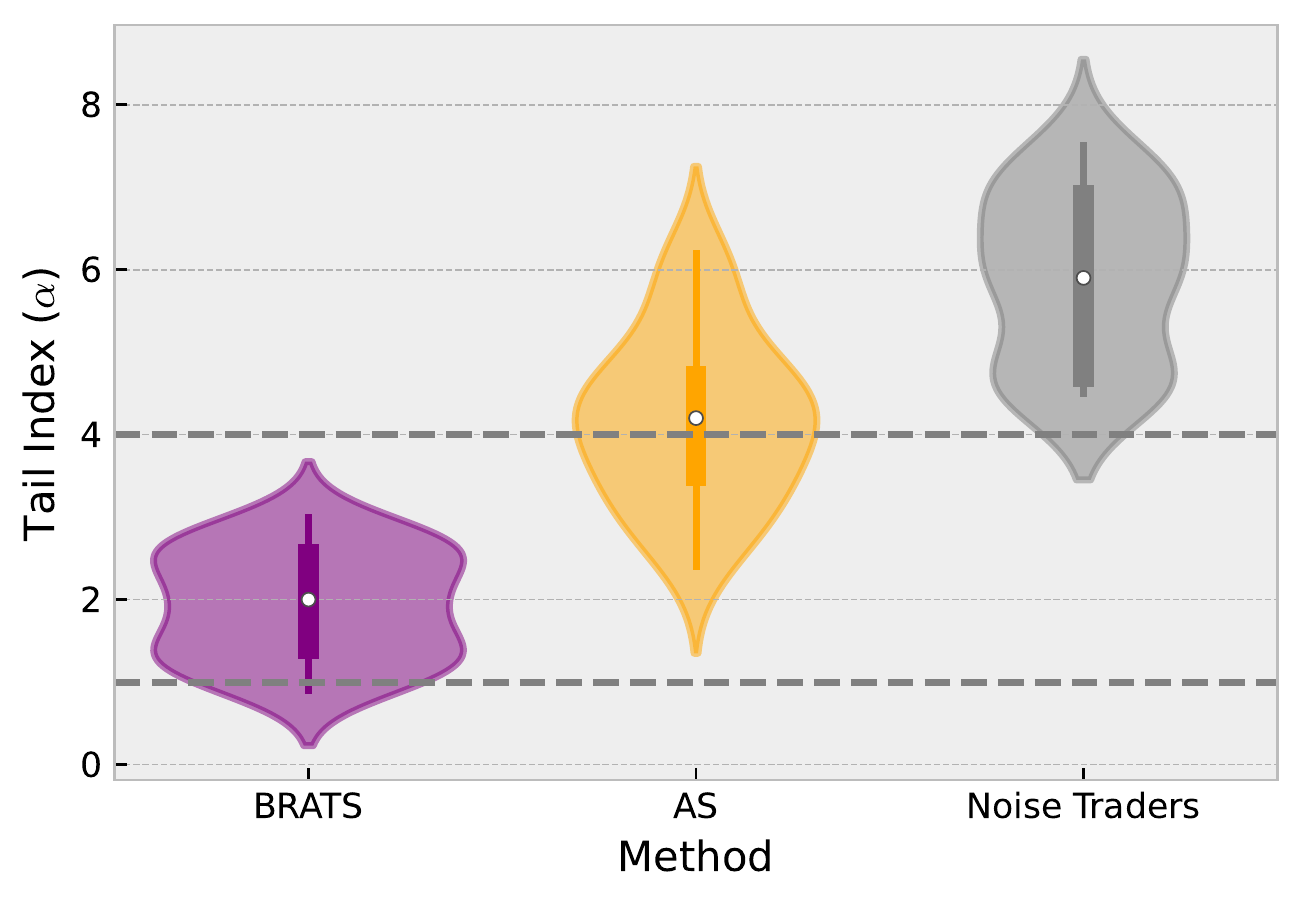}
    \caption{Tail-index estimates $\alpha$ computed using the Hill Estimator. The average values across tail sizes and attendance capacities are used to compute the density for the violin plots (cf. \cref{tblTailIndex}). A lower $\alpha$ indicates a slower decay, and thus a fatter tail. The actual range of tail indices from economic markets is shown with the dashed grey lines.
    }
    \label{figViolin}
\end{figure}

Under the proposed BRATS approach to model market entrance, large abrupt changes in attendance occur more than one would expect from a normal distribution, as displayed in \cref{tblAttendanceRanges}. With the AS model, fatter tails than expected from the normal distribution are also observed, but to a lesser extent. In contrast, the market dynamics generated by agents who follow a random walk (``noise traders") fall entirely in line with the normal distribution. Therefore, the proposed approach demonstrates the ability to recreate an important stylised fact: fat-tailed distributions in terms of (abrupt) resource allocation changes.

We now estimate the tail-index $\alpha$ with a range of common tail sizes, $2.5\%, 5\%, 10\%$~\footnote{We use several tail sizes to account for estimates of $\alpha$ frequently depending on the tail size selection} for (i) the proposed approach, (ii) the canonical AS solution, and (iii) ``noise traders''. To re-iterate, a lower tail-index $\alpha$ implies a heavier tail. The collated results are visualised in \cref{figViolin}, and specific breakdowns are shown in \cref{tblTailIndex}. The tail-indices ($\alpha$'s) produced by the proposed approach consistently fall in the $1-3$ range, as expected by crisis dynamics of emerging ($1 < \alpha < 2$) and established ($2 < \alpha < 4$) markets~\cite{lux2000volatility,lebaron2005extreme}. In contrast, the AS approach often generates larger $\alpha$'s (thinner tails) with approximately half of the estimates falling outside the actual market range for tail indices ($\alpha > 4$). With noise traders, the majority of the $\alpha$'s are found to be well outside this range, indicating even smaller tails.

These findings demonstrate that, unlike its alternatives, the interacting BRATS agents can consistently generate endogenous crises ($3^{+}\sigma$ events), in accordance with observed market crises (indicated by the fat-tails that decay at a realistic rate), while simultaneously improving upon the average convergence.


\subsection{Clustered Volatility}
Having analysed crisis emergence in terms of fat-tails in attendance changes, we turn our attention to the temporal correlations in ``local" volatility, i.e., in the absolute percentage changes in attendance at each time step.

\begin{figure}[!htb]
     \centering
     \begin{subfigure}[b]{0.3\textwidth}
         \centering
         \includegraphics[width=\textwidth]{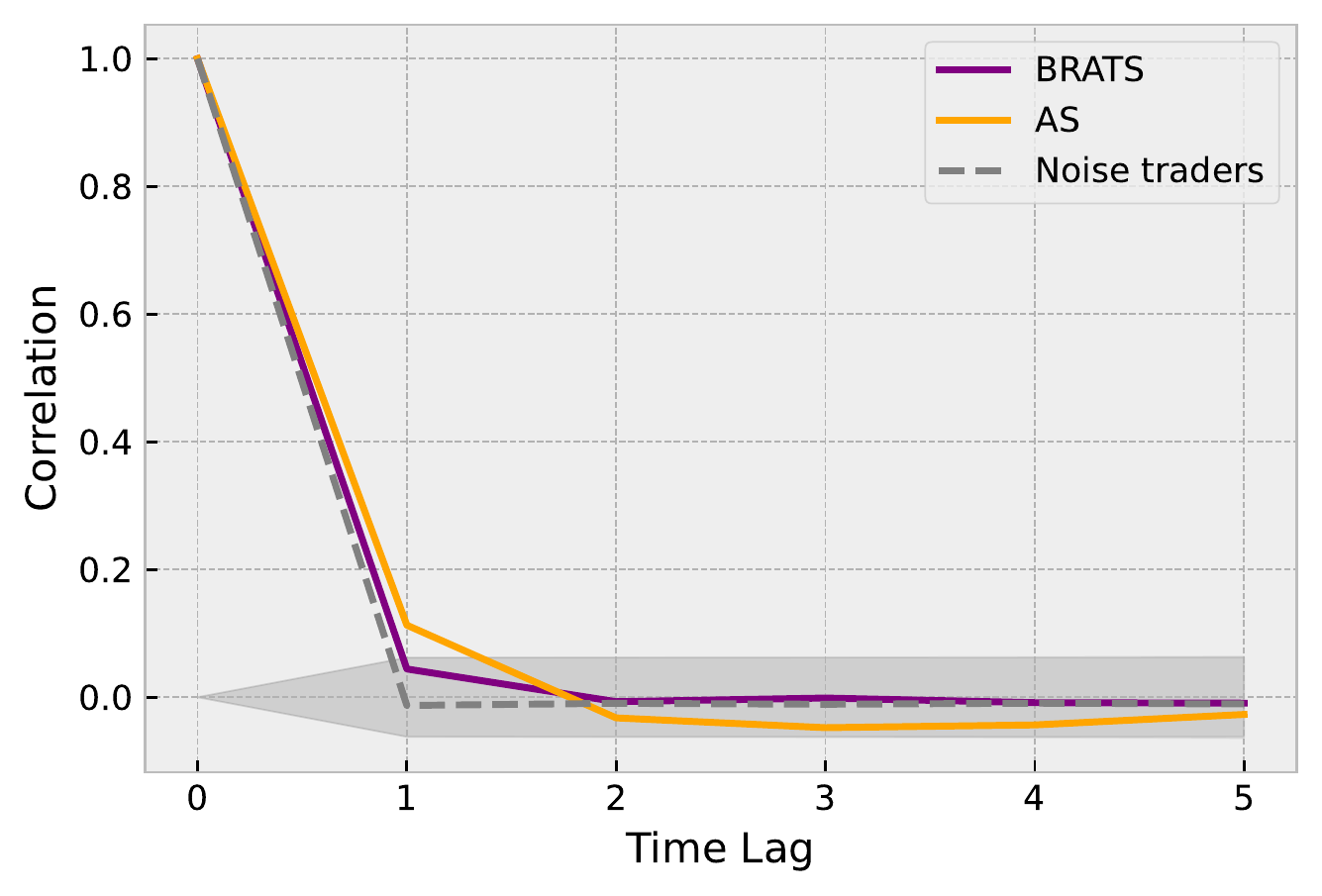}
         \caption{$c=0.1$}
     \end{subfigure}
     \hfill
     \begin{subfigure}[b]{0.3\textwidth}
         \centering
         \includegraphics[width=\textwidth]{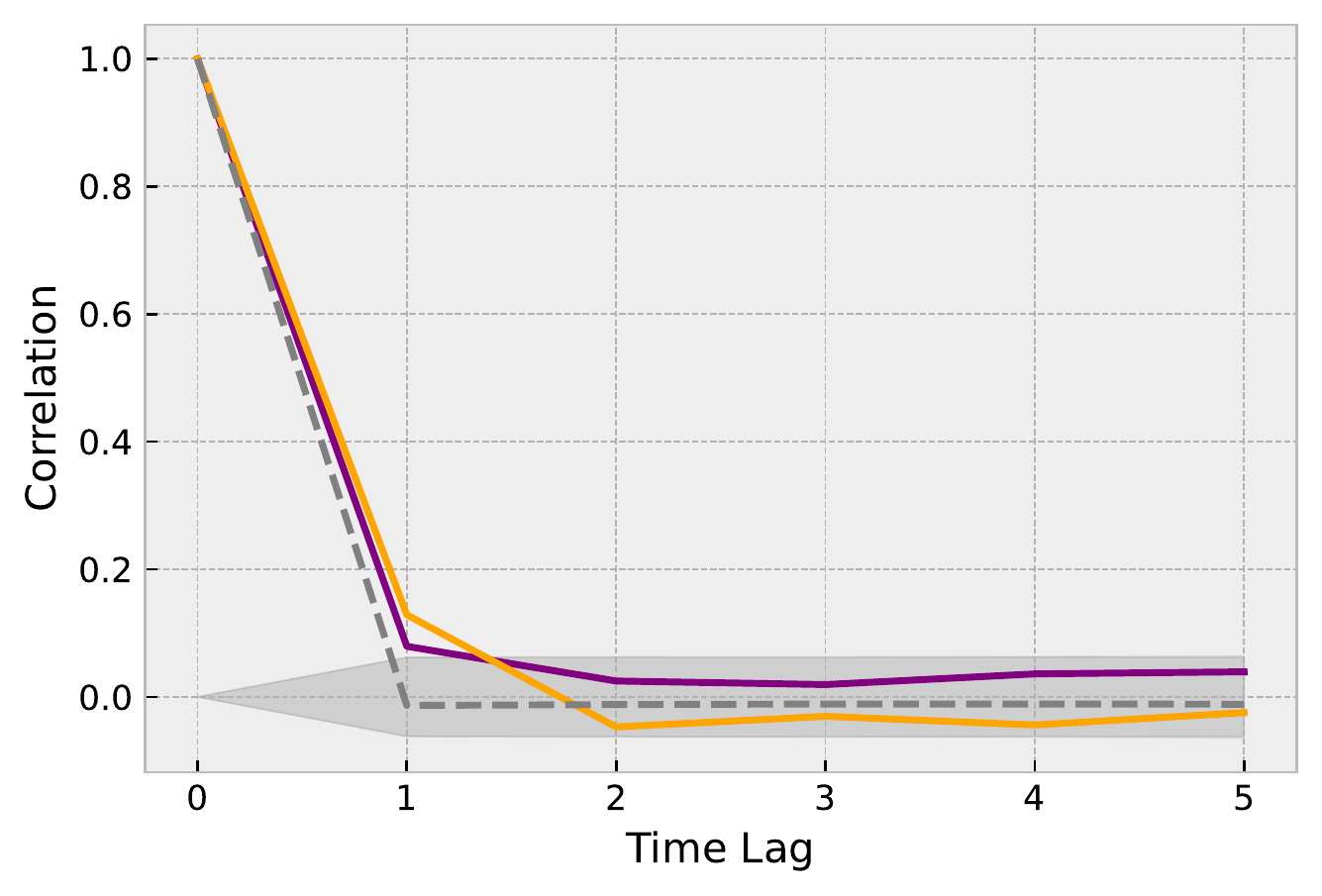}
         \caption{$c=0.2$}
     \end{subfigure}
     \hfill
     \begin{subfigure}[b]{0.3\textwidth}
         \centering
         \includegraphics[width=\textwidth]{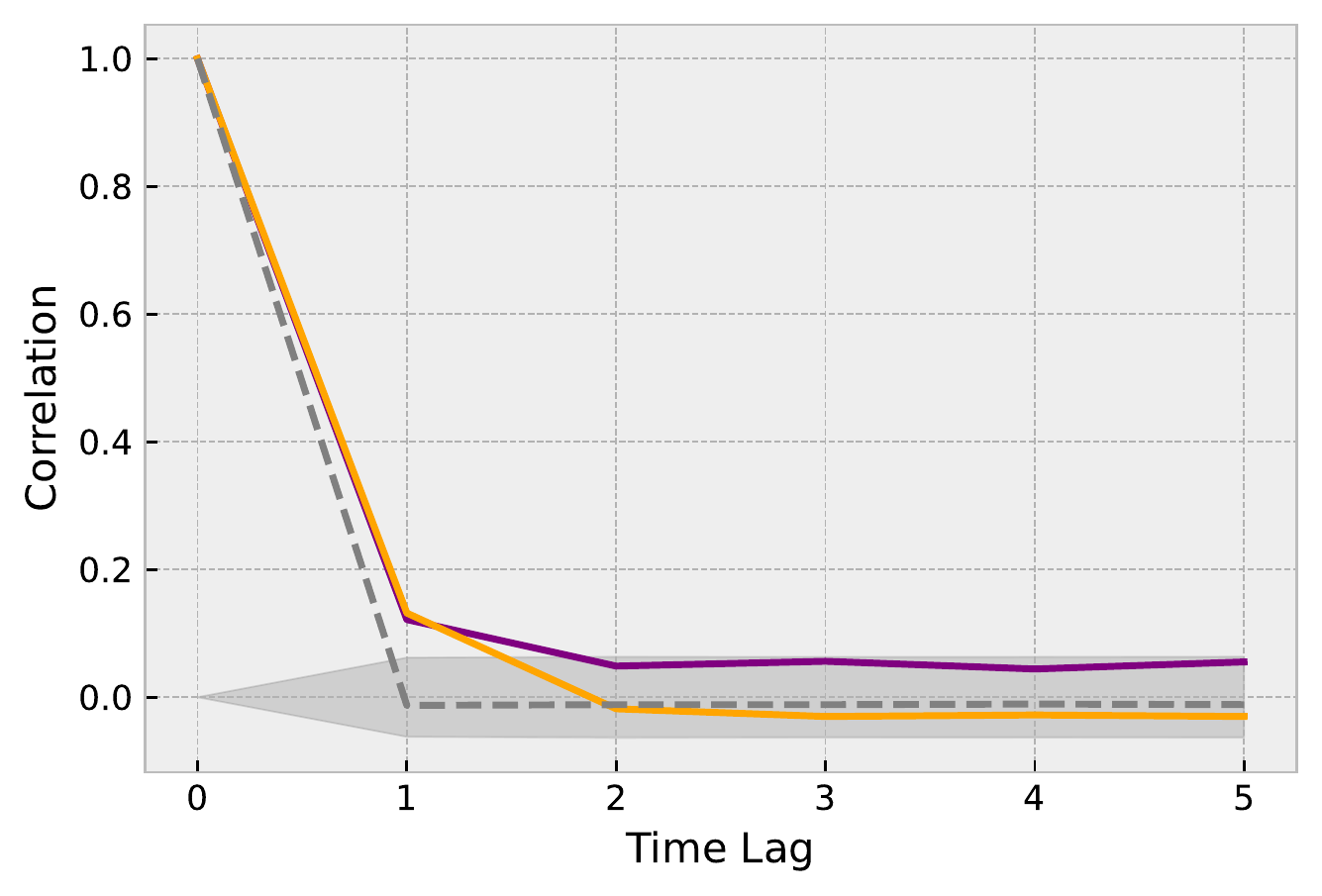}
         \caption{$c=0.3$}
     \end{subfigure}
     \hfill
          \begin{subfigure}[b]{0.3\textwidth}
         \centering
         \includegraphics[width=\textwidth]{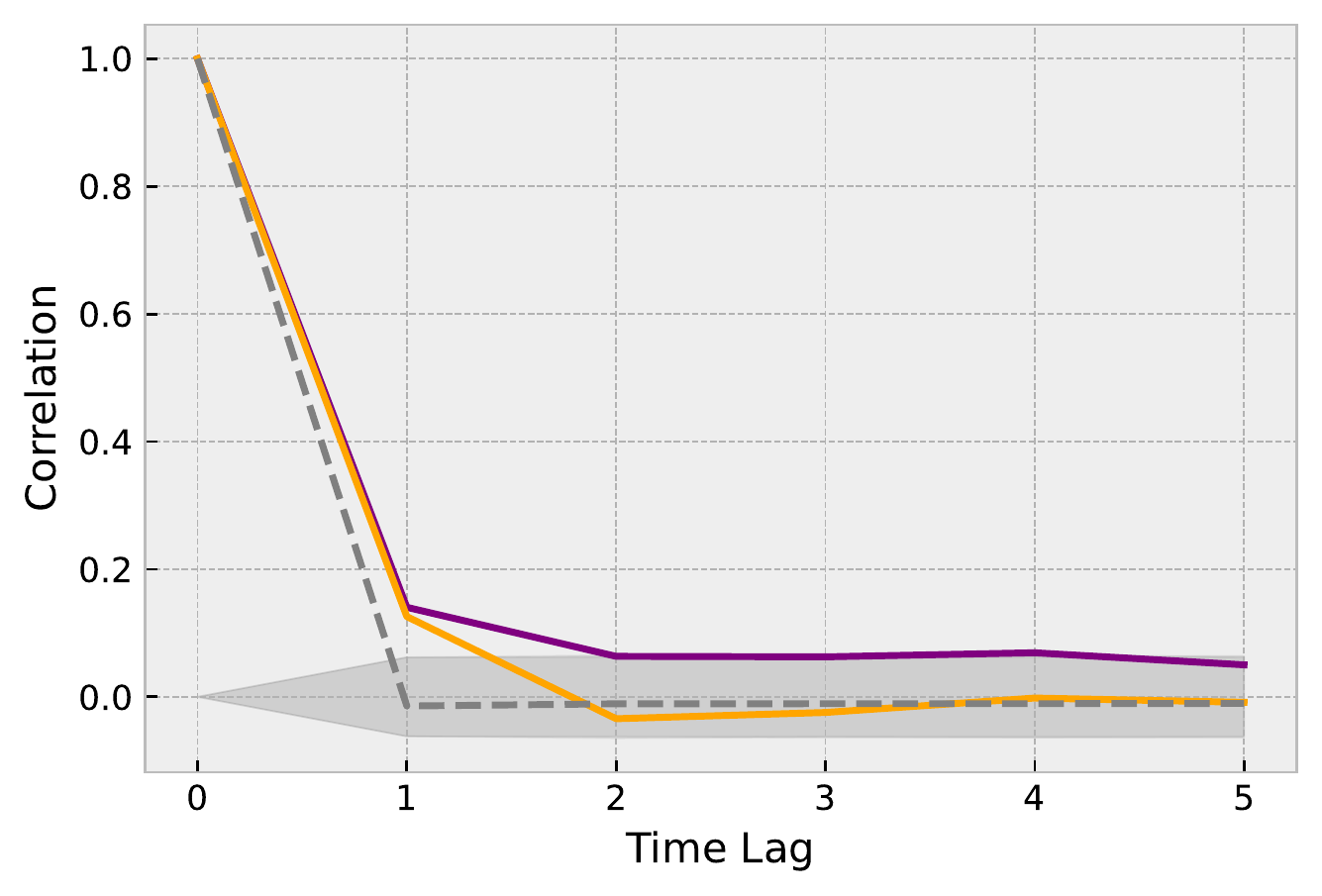}
         \caption{$c=0.4$}
     \end{subfigure}
     \hfill
     \begin{subfigure}[b]{0.3\textwidth}
         \centering
         \includegraphics[width=\textwidth]{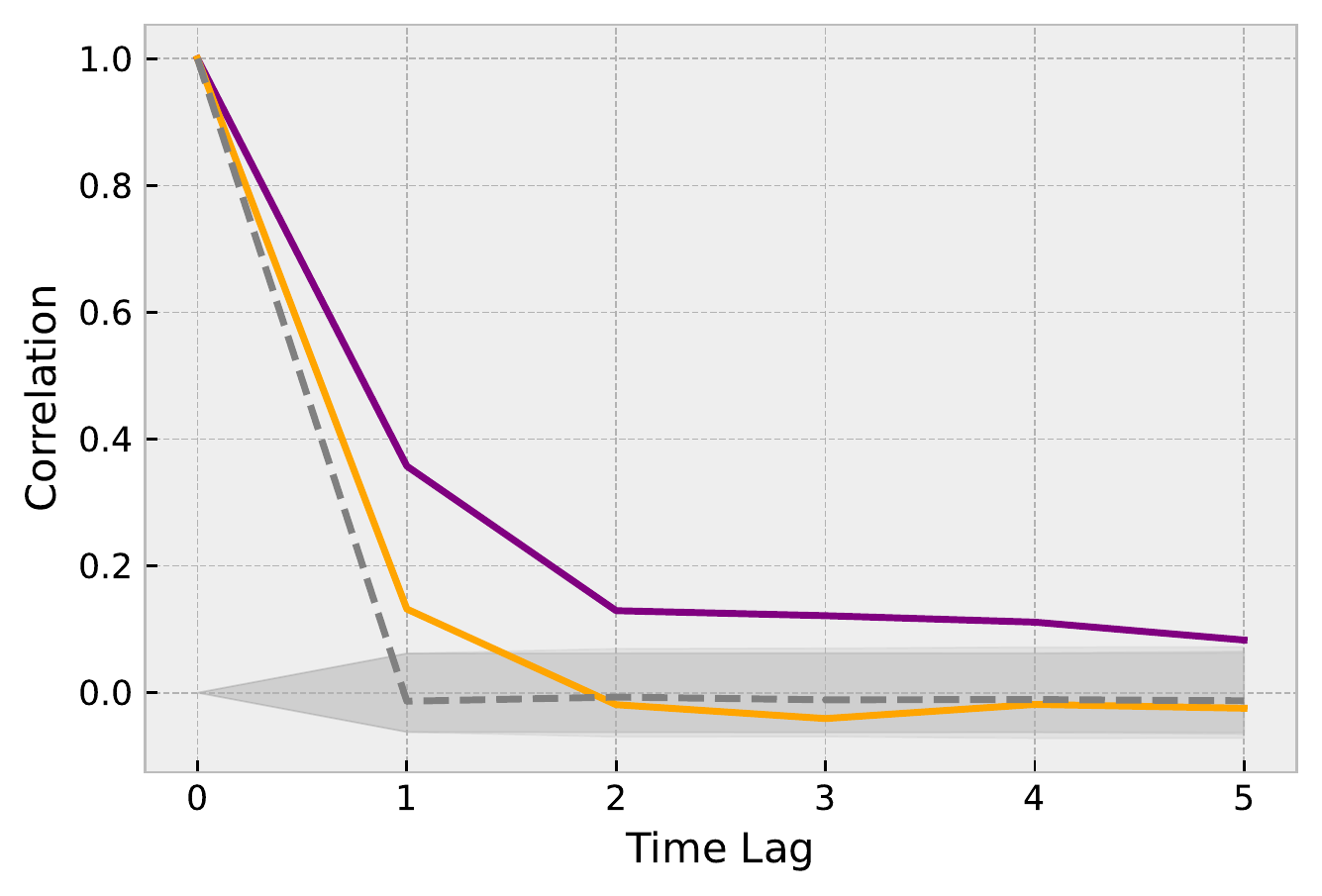}
         \caption{$c=0.5$}
     \end{subfigure}
     \hfill
     \begin{subfigure}[b]{0.3\textwidth}
         \centering
         \includegraphics[width=\textwidth]{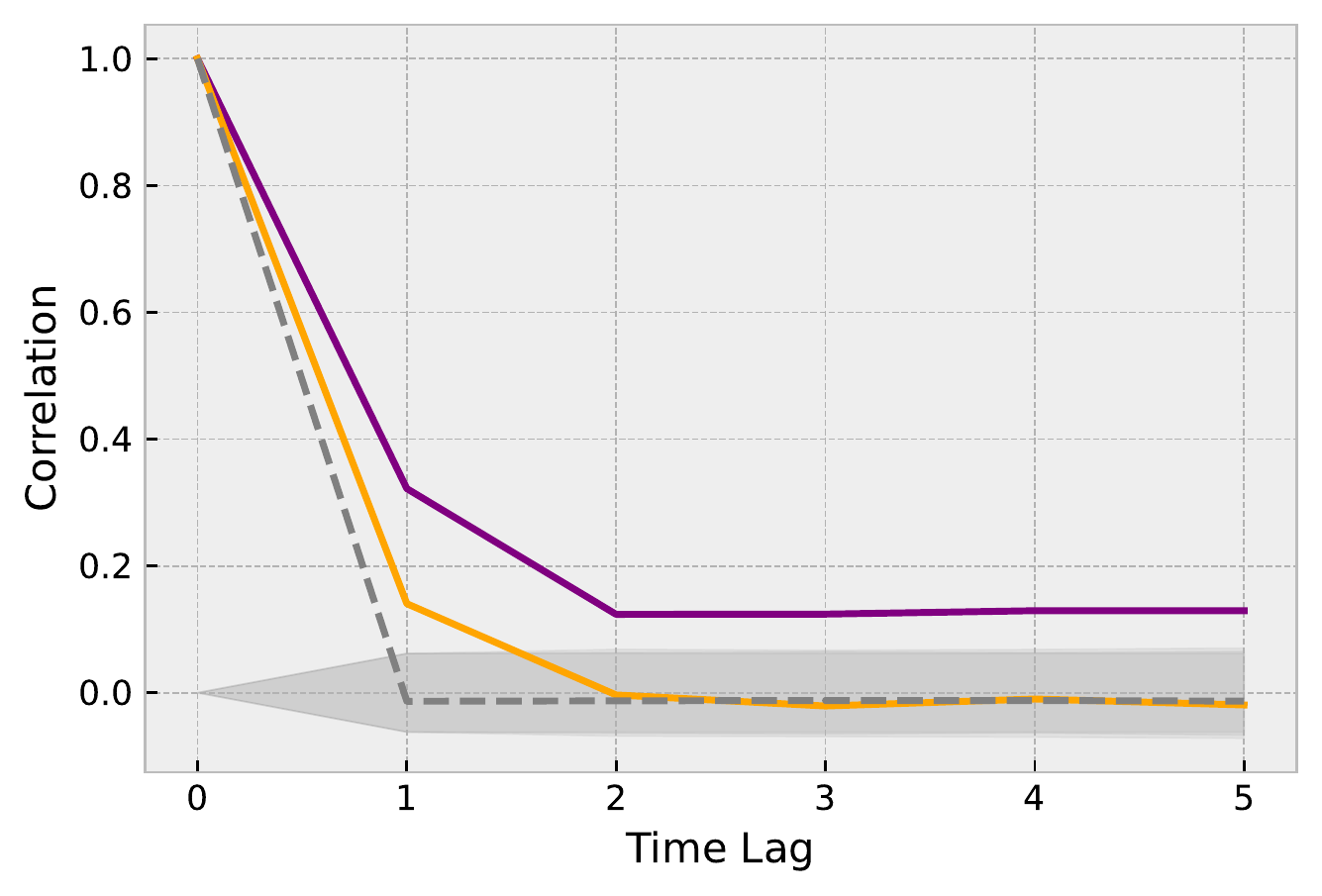}
         \caption{$c=0.6$}\label{figClusteredVolatility0.6}
     \end{subfigure}
     \hfill
          \begin{subfigure}[b]{0.3\textwidth}
         \centering
         \includegraphics[width=\textwidth]{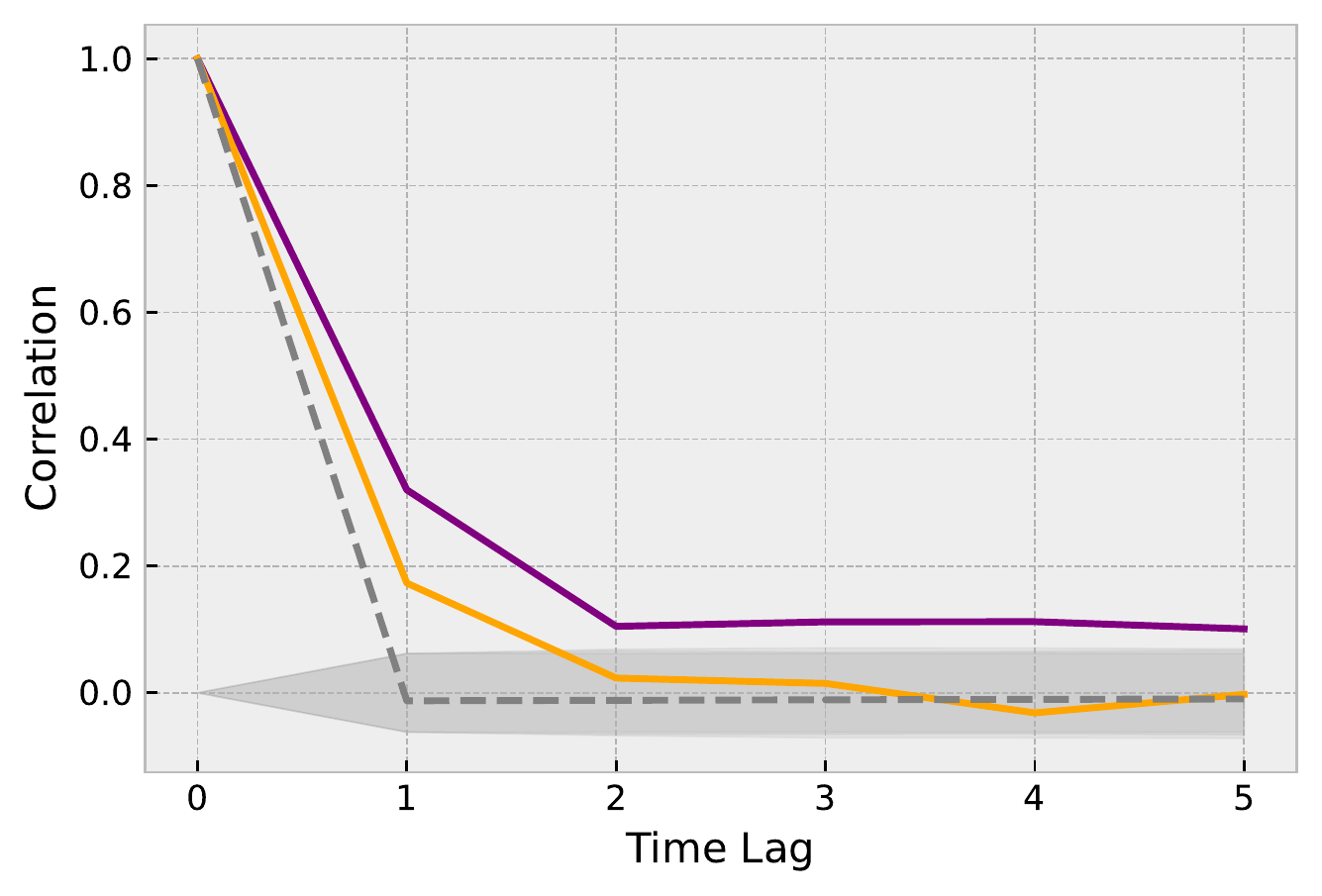}
         \caption{$c=0.7$}\label{figClusteredVolatility0.7}
     \end{subfigure}
     \hfill
     \begin{subfigure}[b]{0.3\textwidth}
         \centering
         \includegraphics[width=\textwidth]{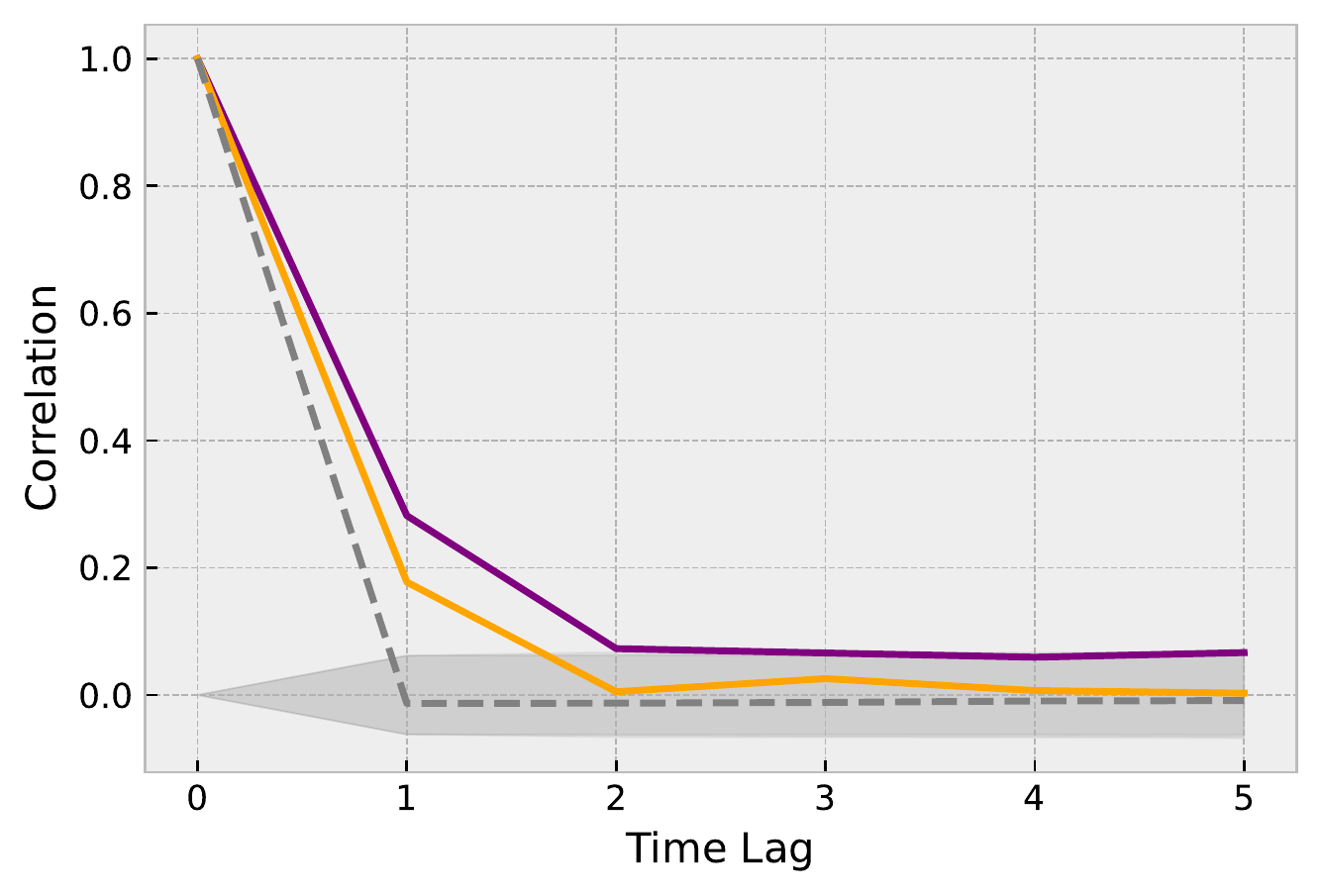}
         \caption{$c=0.8$}
     \end{subfigure}
     \hfill
     \begin{subfigure}[b]{0.3\textwidth}
         \centering
         \includegraphics[width=\textwidth]{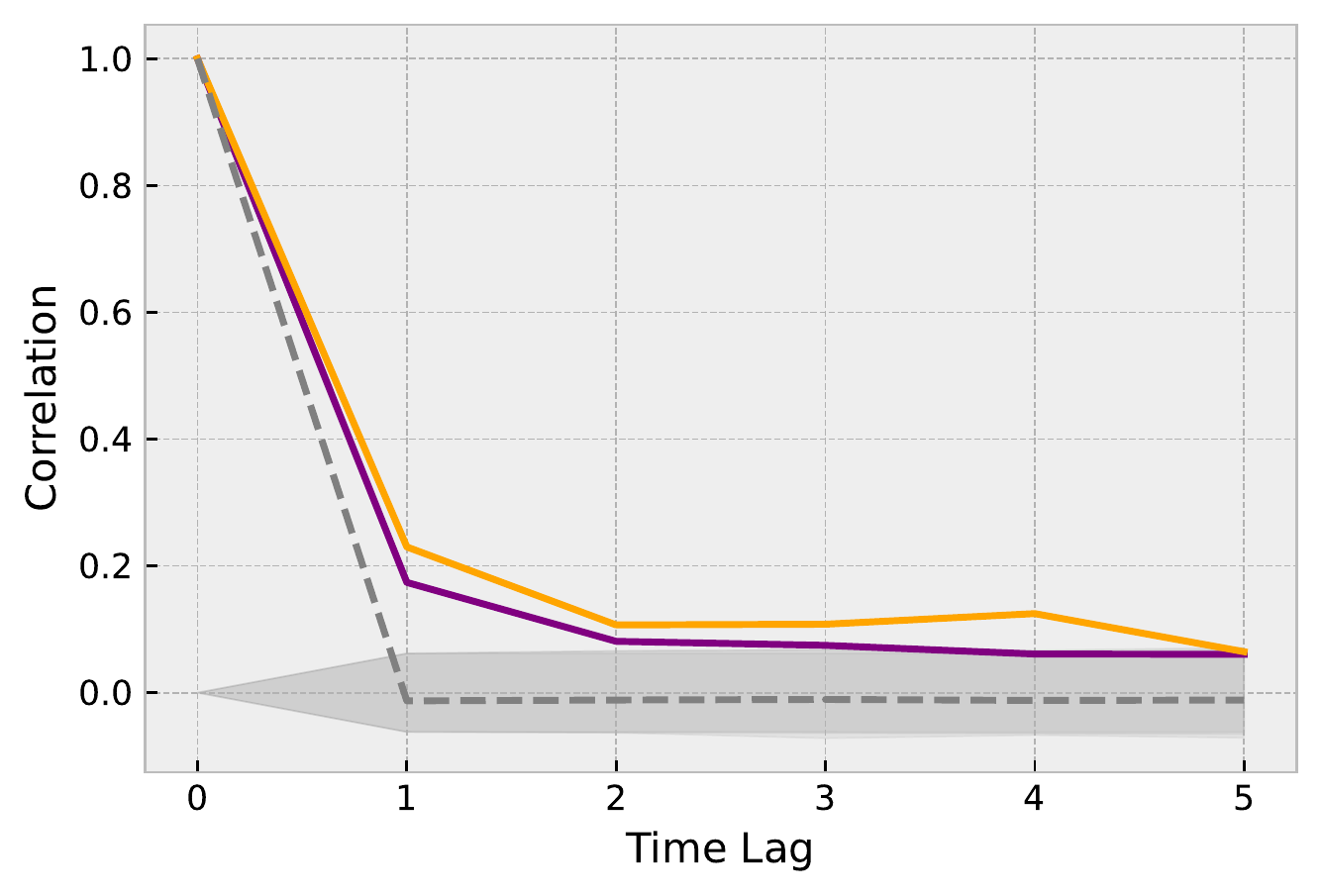}
         \caption{$c=0.9$}
     \end{subfigure}
        \caption{Autocorrelation in the attendance volatility, showing the presence of clustered volatility. Purple (orange) represents the proposed BRATS model (adaptive strategies). The dashed grey line represents noise traders. The filled grey region around $0$ highlights the area of no statistically significant correlations.}
        \label{figClusteredVolatility}
\end{figure}

The temporal volatility correlations in the market entrance game are presented in \cref{figClusteredVolatility}. The  BRATS approach shows statistically significant autocorrelations for up to five time steps, which is particularly notable in \cref{figClusteredVolatility0.6}. These results are in contrast to the results produced by the AS approach, which generally shows fewer temporal correlations in volatility, and in fact, typically displays no significant temporal correlation beyond one time step (e.g. \cref{figClusteredVolatility0.7}). These findings indicate that the ability to recreate the clustered volatility observed in actual markets across multiple temporal lags is not present in the canonical AS model. Furthermore, the noise traders show no temporal volatility correlation (beyond lag zero). The results shown in \cref{figClusteredVolatility} confirm that volatility clustering can arise purely endogenously, as the observed correlations occurred without any external changes to the system. That is not to say that volatility does not also arise due to external news, but rather that in addition, periods of increased volatility can arise purely endogenously in the absence of such news.

To explain the endogenous emergence of periods of increased volatility, we analyse whether the heterogeneity (diversity) of agents' beliefs can be seen as a leading indicator of the future volatility change. In doing so, we employ a diversity measure described in \cref{appendixDiversity}, quantified by the normalised entropy of the agent population, suitably adopted for BRATS and AS approaches. When the diversity of the population decreases, this can indicate a group or herd formation, and thus, a reduction in heterogeneity of the population.

\begin{figure}[!htb]
    \centering
    \begin{subfigure}[b]{0.3\textwidth}
    \includegraphics[width=\textwidth]{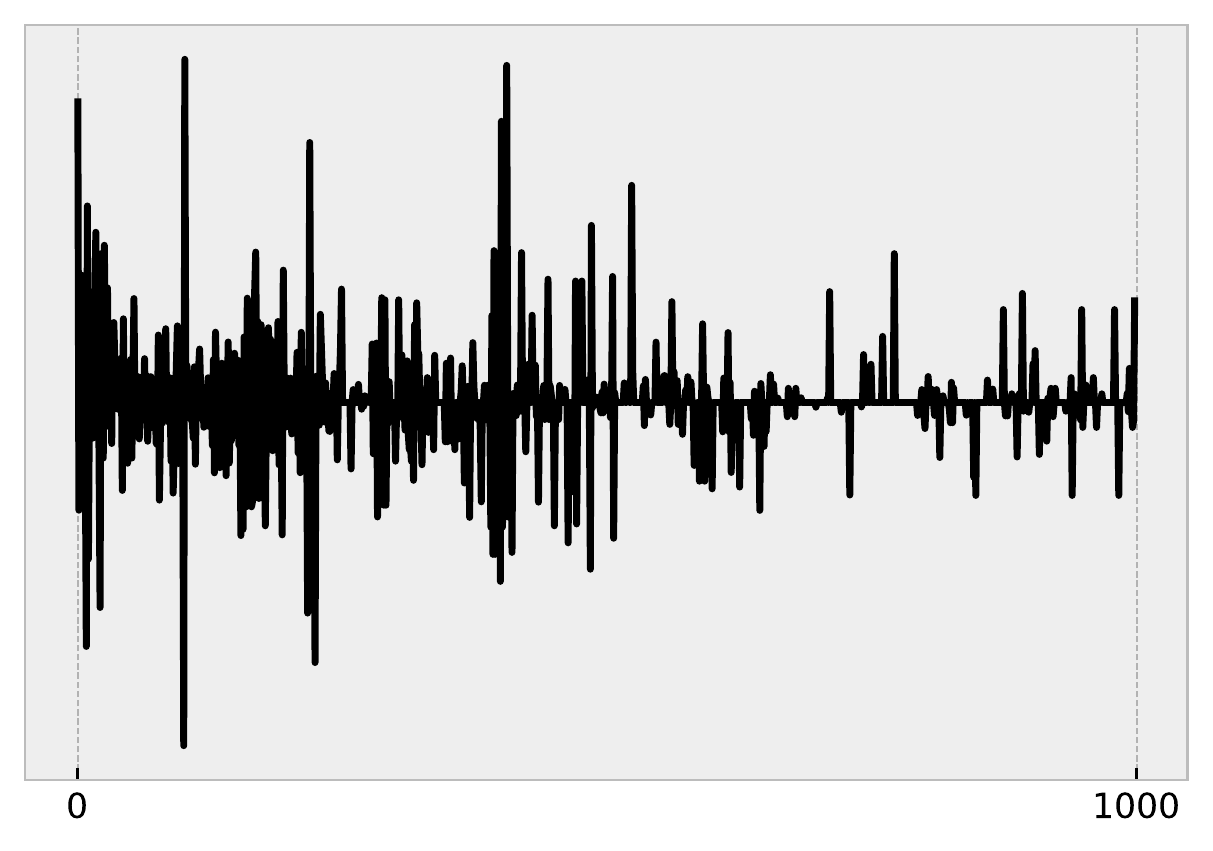}
    \caption{Diversity Change}
    \end{subfigure}
    \begin{subfigure}[b]{0.3\textwidth}
    \includegraphics[width=\textwidth]{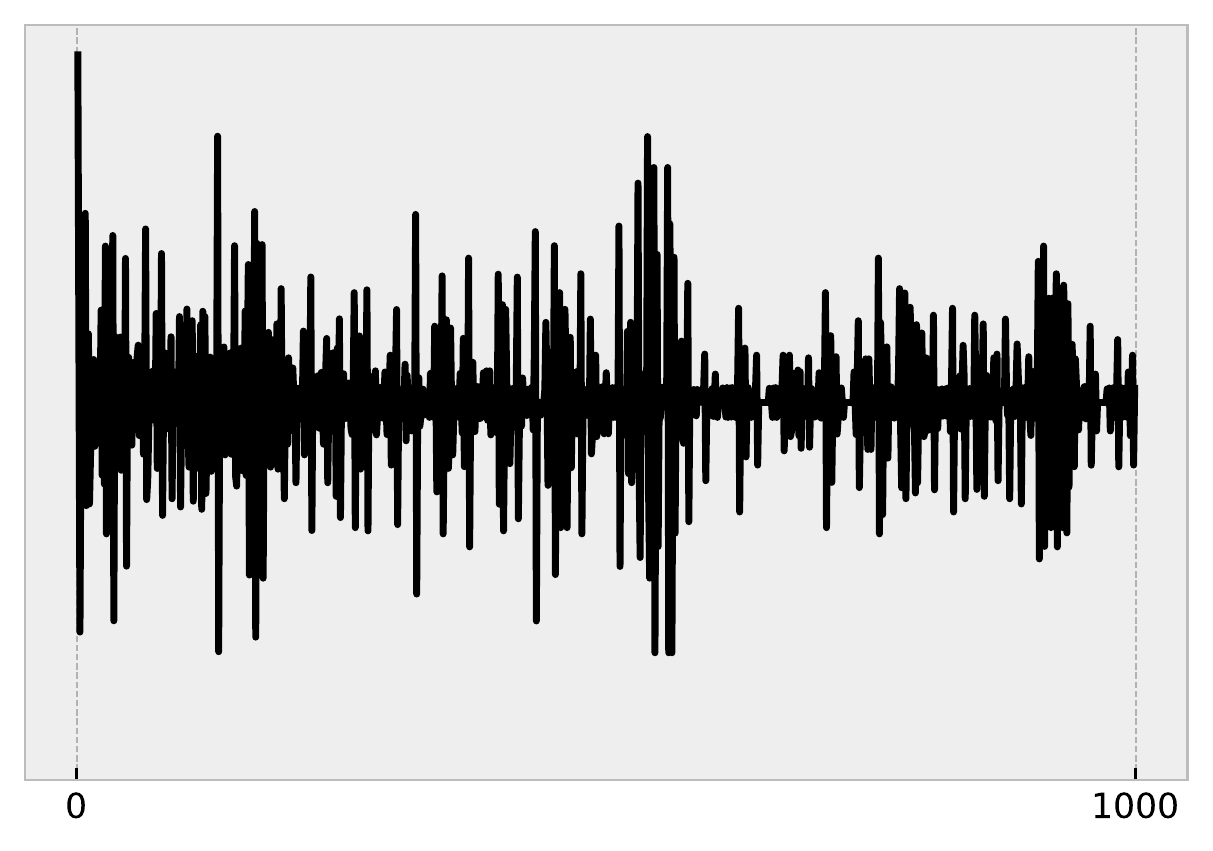}
    \caption{Volatility Change}
    \end{subfigure}
    \caption{Example stationary transformations from one realisation}
    \label{figExampleTimeSeries}
\end{figure}

\begin{figure}[!htb]
     \centering
     \begin{subfigure}[b]{0.3\textwidth}
         \centering
         \includegraphics[width=\textwidth]{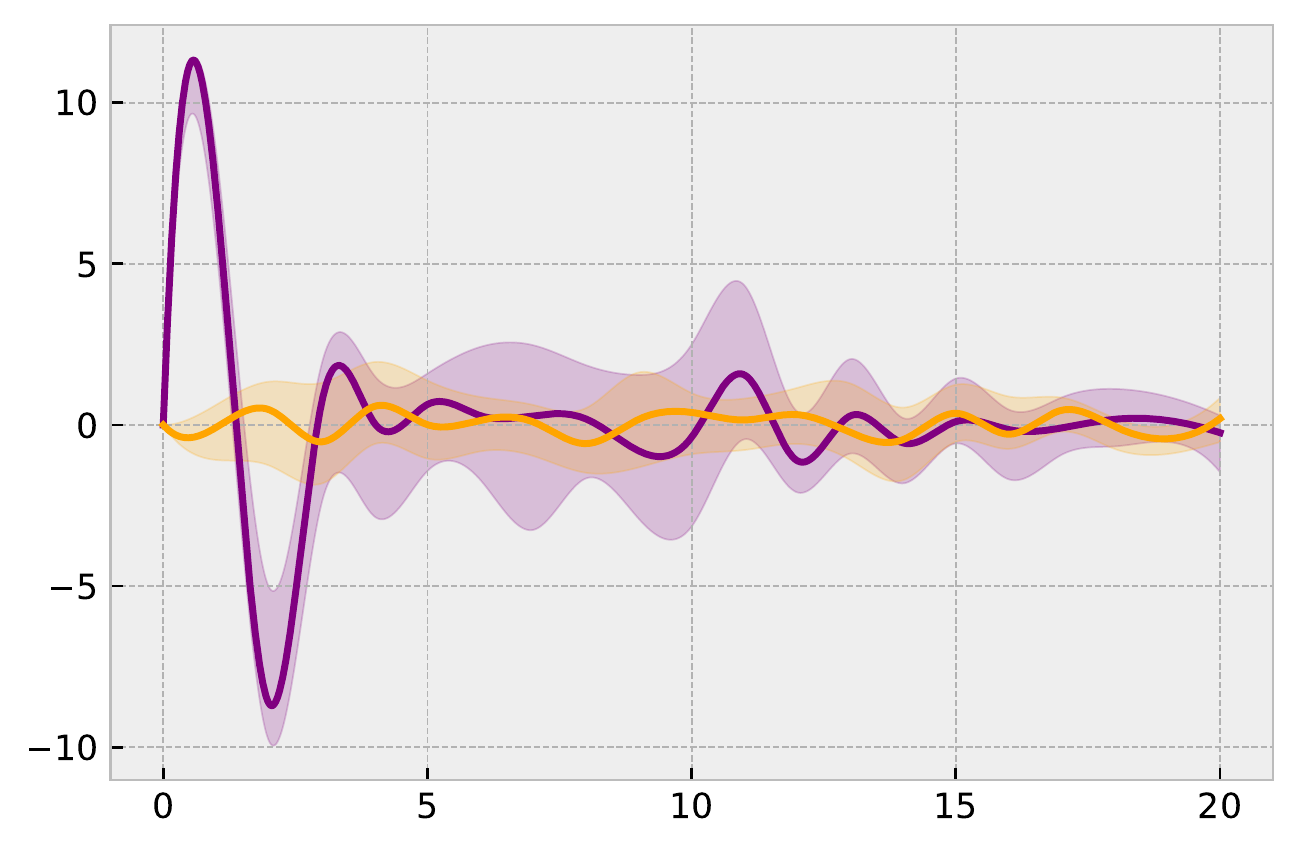}
         \caption{$c=0.1$}
     \end{subfigure}
     \hfill
     \begin{subfigure}[b]{0.3\textwidth}
         \centering
         \includegraphics[width=\textwidth]{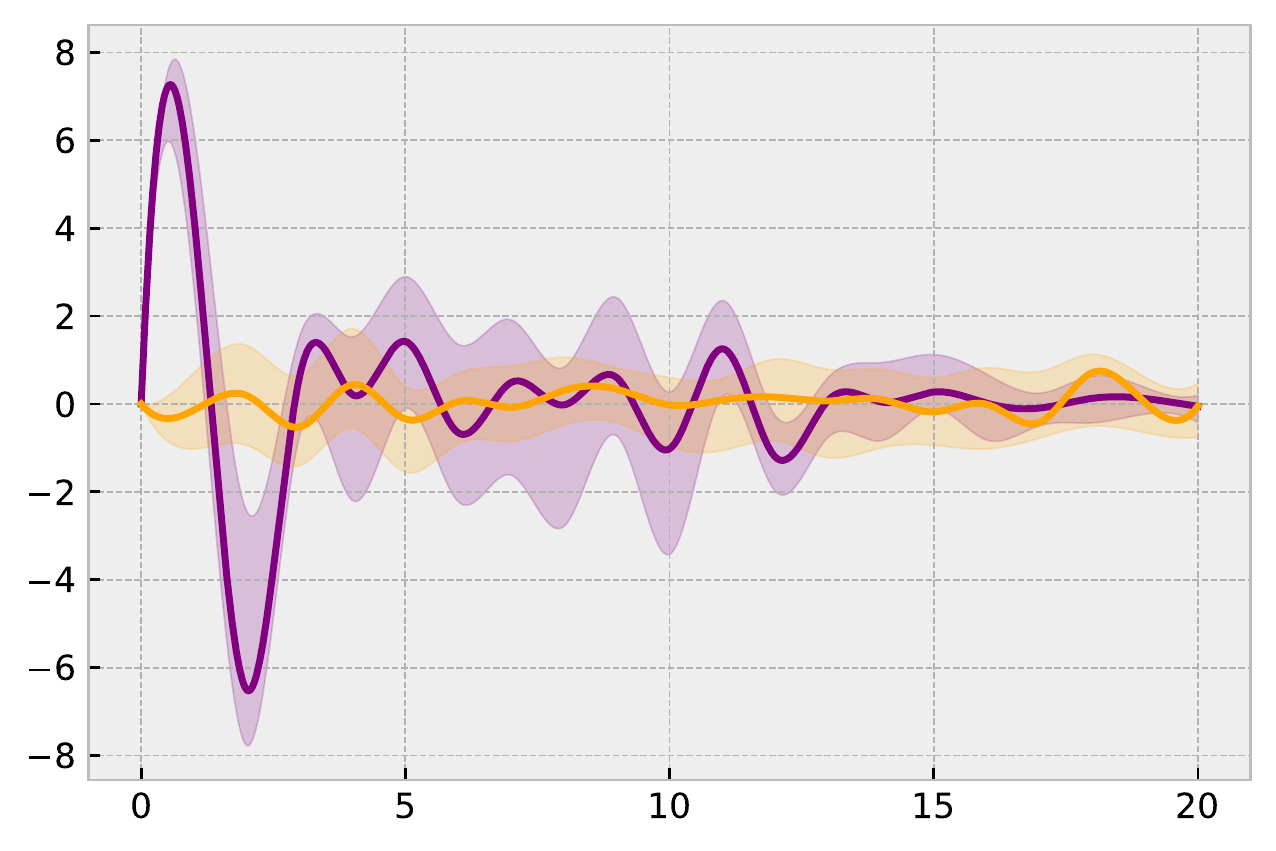}
         \caption{$c=0.2$}
     \end{subfigure}
     \hfill
     \begin{subfigure}[b]{0.3\textwidth}
         \centering
         \includegraphics[width=\textwidth]{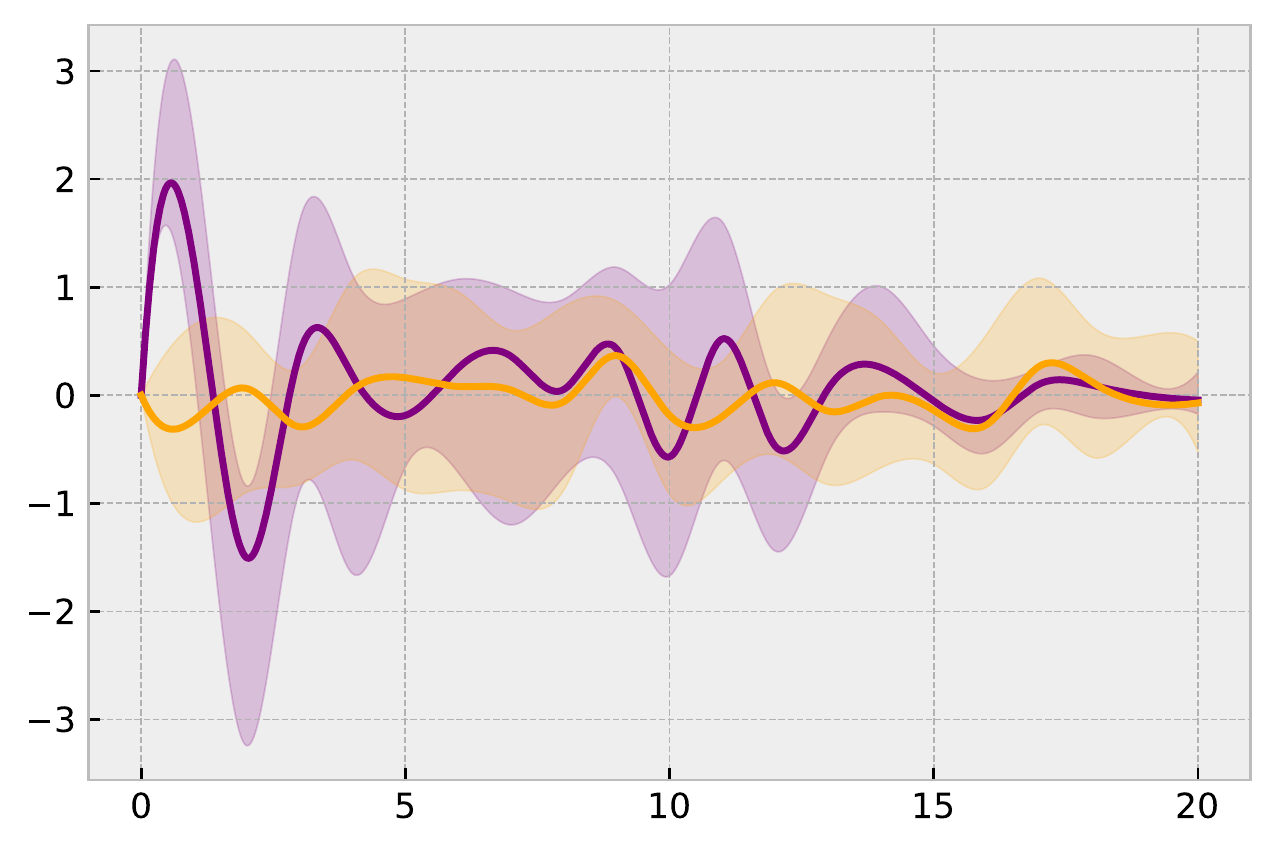}
         \caption{$c=0.3$}
     \end{subfigure}
     \hfill
          \begin{subfigure}[b]{0.3\textwidth}
         \centering
         \includegraphics[width=\textwidth]{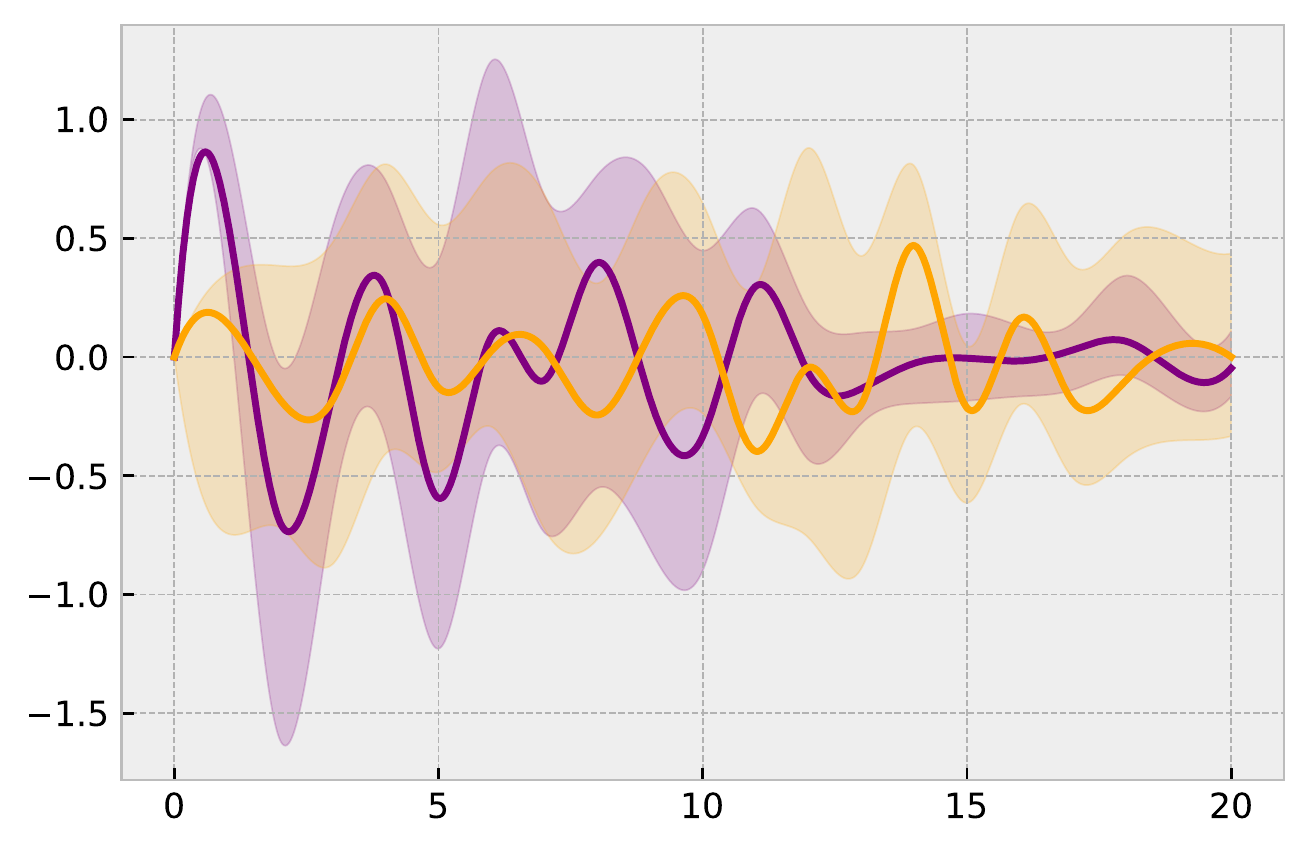}
         \caption{$c=0.4$}
     \end{subfigure}
     \hfill
     \begin{subfigure}[b]{0.3\textwidth}
         \centering
         \includegraphics[width=\textwidth]{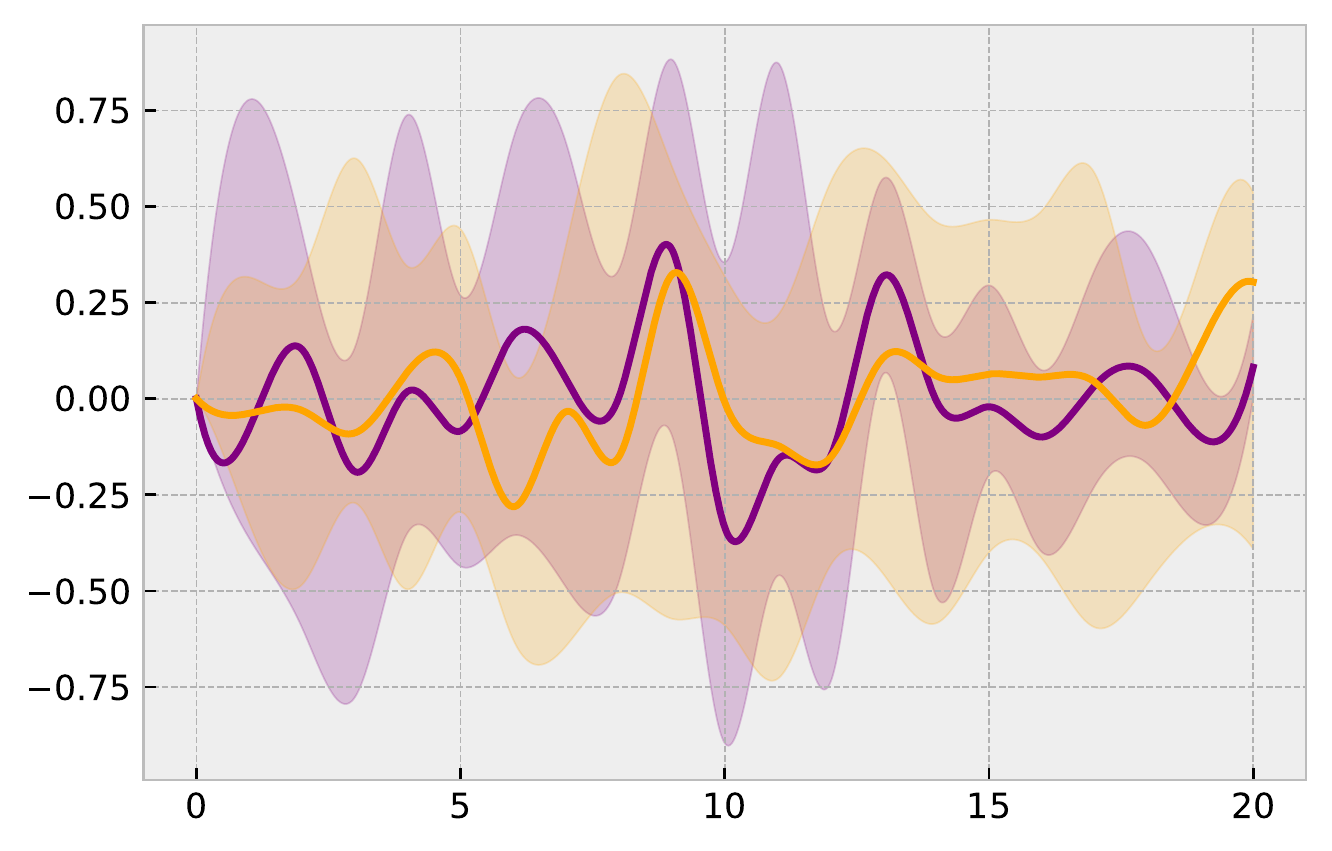}
         \caption{$c=0.5$}
     \end{subfigure}
     \hfill
     \begin{subfigure}[b]{0.3\textwidth}
         \centering
         \includegraphics[width=\textwidth]{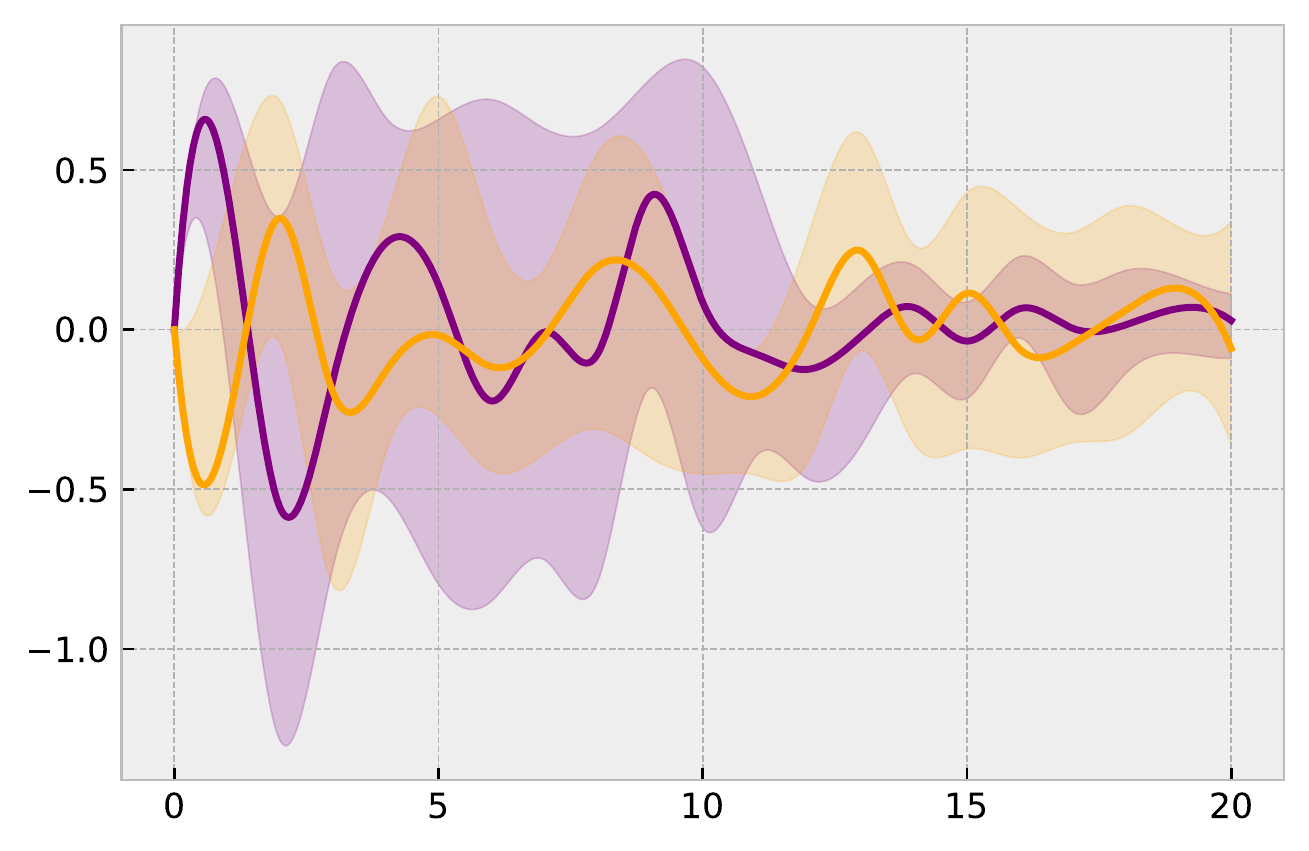}
         \caption{$c=0.6$}
     \end{subfigure}
     \hfill
          \begin{subfigure}[b]{0.3\textwidth}
         \centering
         \includegraphics[width=\textwidth]{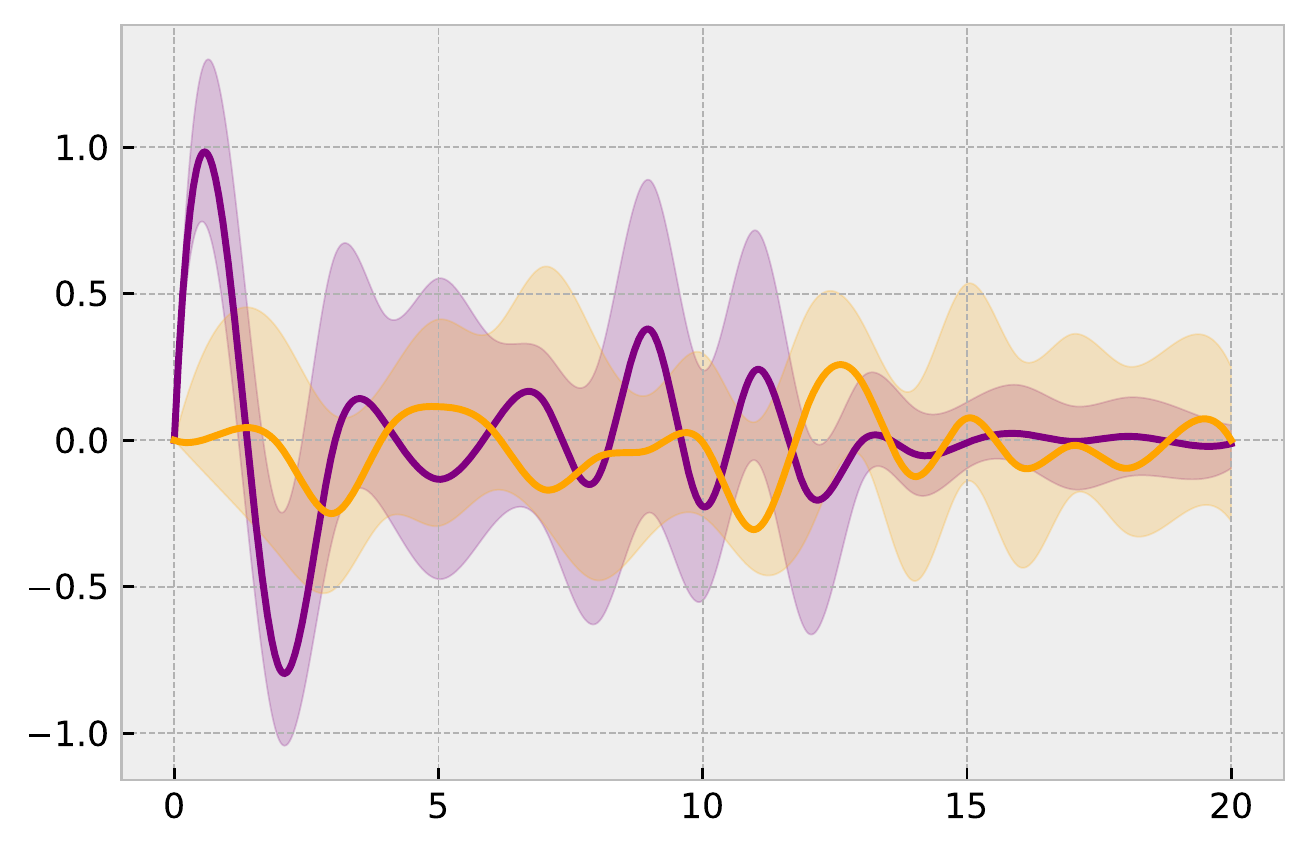}
         \caption{$c=0.7$}
     \end{subfigure}
     \hfill
     \begin{subfigure}[b]{0.3\textwidth}
         \centering
         \includegraphics[width=\textwidth]{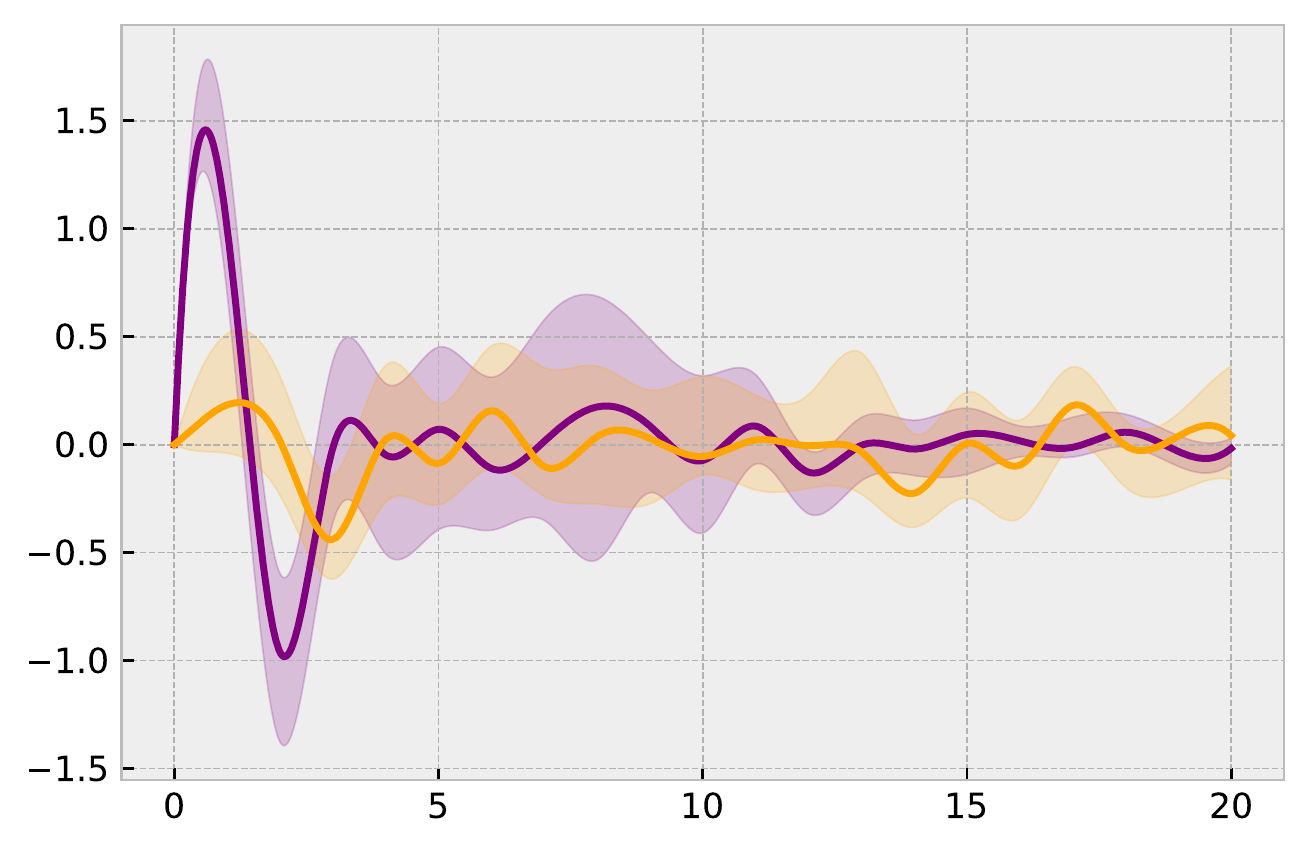}
         \caption{$c=0.8$}
     \end{subfigure}
     \hfill
     \begin{subfigure}[b]{0.3\textwidth}
         \centering
         \includegraphics[width=\textwidth]{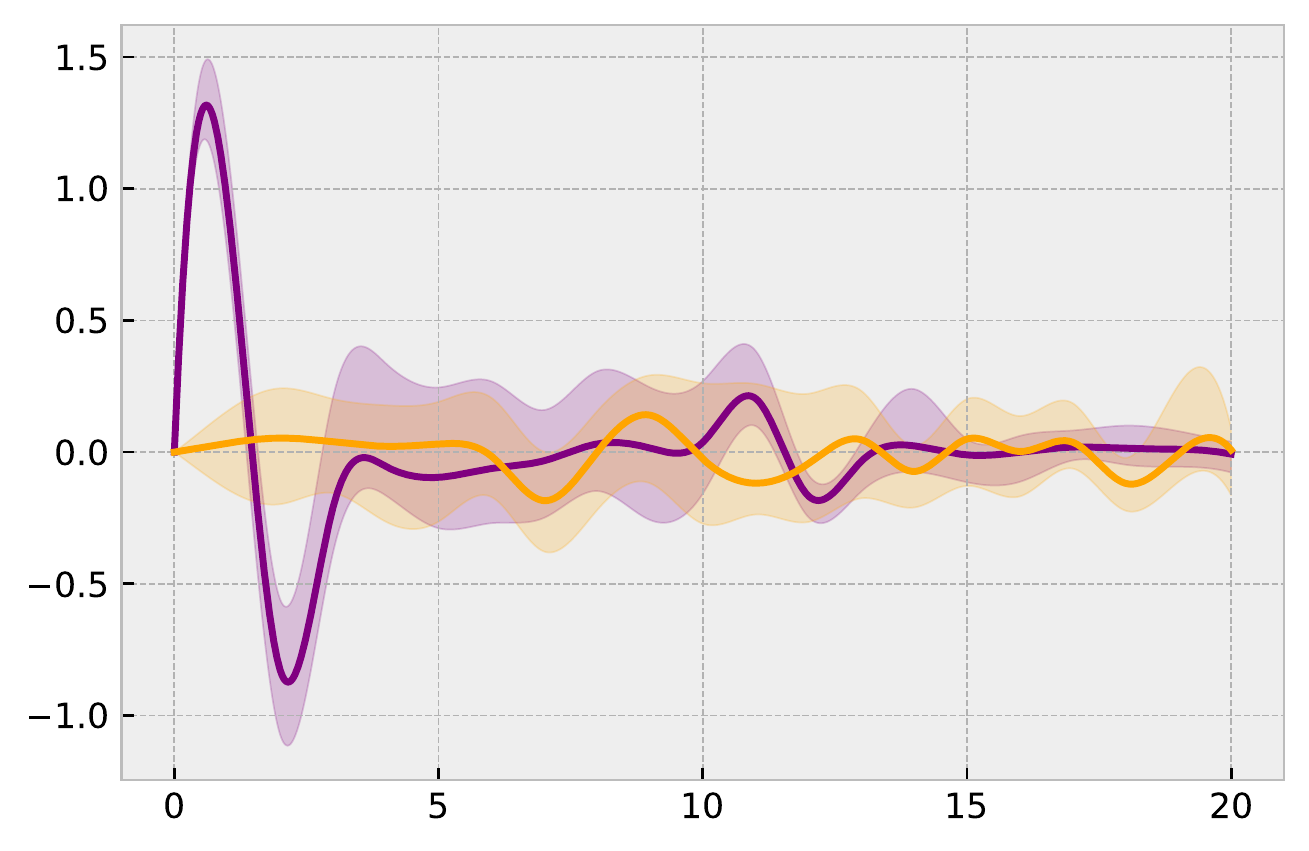}
         \caption{$c=0.9$}
     \end{subfigure}
        \caption{Smooth Impulse Response Functions. Purple (orange) represents the proposed BRATS model (adaptive strategies). The dark-line represents the median from across runs, and the shaded lighter region the interquartile range. The $y$-axis shows the change in volatility difference following the impulse, and the $x$-axis shows the time since the impulse.
        }
        \label{figImpulseResponse}
\end{figure}

We find that changes in the population diversity  are predictive of (or Granger causing) changes in attendance (details are provided in \cref{appendixGrangerProcess}). An example of a realisation of these two time series is presented in \cref{figExampleTimeSeries}. Crucially, these findings show that the distribution of agent beliefs is a significant predictor of volatility. To measure the effect that the diversity change has on the resulting volatility, beyond merely being predictive of this change, we use an impulse-response function  quantifying the impact on volatility generated by an ``impulse" of diversity (\cref{figImpulseResponse}). One unit shock in diversity change results in an initial consistent increase in positive volatility change, followed by a ``correction" to a negative change in volatility. This correction is then followed by a destabilising period of oscillatory dynamics indicating larger changes and uncertainty in the volatility. This larger uncertainty is  marked by the larger interquartile range. Finally, the dynamics settle down after approximately $10$ periods, without a discernible influence beyond this interval. 

When the diversity of the population is low, the system is in a ``turbulent" regime \cite{schmitt2017herding} where volatility is high due to the interdependence of agent decisions. Over time, agents react and modify their beliefs. Due to the agents’ alternate learning updates, the belief correlation declines, and the system returns to a calm regime with less volatility, characterised by higher heterogeneity of the population. This reduction in the lagged auto-correlation in volatility is demonstrated in \cref{figClusteredVolatility}. The reduction is also reflected in the oscillatory period in \cref{figImpulseResponse} during which the impulse response flattens over time. 
In contrast, the results produced by the AS 
do not provide clear outcomes in terms of impulse-response, as shown in  \cref{figImpulseResponse}. This indicates that clustered volatility cannot be explained, or predicted, based upon the AS approach.

\section{Discussion} 

The formation of crises in economic markets is a well-established phenomenon, dating back to at least the Dutch ``tulip mania" of 1636 which abruptly burst overnight \cite{brunnermeier2016bubbles,garber1989tulipmania}. This boom and bust are often explained due to the irrationality of market participants \cite{mackay2012extraordinary}. The importance of  this irrationality was discussed by Keynes who explained price fluctuations with the beauty contest game~\cite{keynes1937general}. Specifically, Keynes pointed out that ``I-think-you-think-they-think...-type of regress" better represented investor behaviour, with pricing driven by  investors' beliefs about other investors, rather than by the asset fundamentals.

While this ``I-think-you-think-they-think...-type of regress" has been identified as a natural representation for agent behaviour in market games \cite{casti1996seeing},  modelling these recursive beliefs becomes problematic becomes problematic due to the potential infinite  reasoning \cite{koppl2002all}.  
A recently introduced approach, based on the Quantal Hierarchy (QH) model \cite{evans2021bounded}, addressed limitations of this higher-order reasoning by ``breaking" at various points of recursion, thus accounting for boundedly rational agents and preventing infinite regress. In this work, exemplified by a concise market configuration, the canonical El Farol Bar Problem, we extended the QH model to capture crisis emergence in market entrance games. Crucially, this analysis showed that bounded strategic reasoning can lead to higher probabilities of crises (i.e., fat-tails in attendance fluctuations), matching the dynamics of actual economic markets. Furthermore, we related the resulting volatility to the diversity of strategic reasoning resources across the agent population. 


Of course, convergence to a desirable market capacity, an ``equilibrium", has been demonstrated for the El Farol bar problem before~\cite{arthur1994inductive,fogel1999inductive,bell2003coordination,challet2004shedding}. Moreover, it has been shown that even a large number ($N \to \infty$) of zero-intelligence agents can self-organise to the desired capacity under relatively general conditions, simply due to the law of large numbers~\cite{challet2004shedding}. The model proposed in this work performed at least as good as these approaches, even in systems with relatively low number of heterogeneous agents ($N=100$), as long as the agents follow boundedly rational strategic reasoning.



However,  convergence to equilibrium is a necessary but not sufficient characteristic for an adequate model of actual markets, in which crises can endogenously form and volatility can cluster in time.
Hence, we systemically explored the formation of crises based on abrupt deviations from an equilibrium state. Specifically, we analysed the tail-indices of  \textit{endogenously} emerging crises, categorised by significant  changes in attendance ($3^{+}\sigma$ events), where the system can be seen as briefly being out-of-equilibrium. This analysis demonstrated that these fat tails are significantly heavier than one would expect under a rational expectations model without the arrival of any external news. We related these resultant tail indices to actual market data, showing that the fat tails produced by the BRATS approach decay at a realistic rate.

Finally, we traced and examined the endogenous emergence of periods of volatility based on changes in agent beliefs, and related this to the agent learning process. A key stylised fact of economic markets --- clustered volatility --- was recreated using a concise learning update of agents' reasoning resources. 
Importantly, the change in the overall diversity of agent beliefs was predictive of the change in future volatility, indicating that the heterogeneity of agent beliefs is a crucial predictor of volatility.

These findings strengthened the conjecture that the diversity of market participant beliefs is one of the ``most important propagation mechanism[s] of economic volatility"~\cite{kurz2001endogenous}. A similar conclusion was reached by~\cite{lux2000volatility}, where periods of instabilities caused by the ratio of chartists and fundamentalists in the market were ``quickly brought to an end by stabilising tendencies".
Yet, the diversity of strategies has not been fully explored in the context of the El Farol bar problem~\cite{rand2007farol}. Here we offered a way to analyse the population diversity in terms of reasoning resources. In doing so, we quantified the market dynamics, confirming that during a period of high volatility, if agent beliefs are correlated, the volatility persists. As the agents attempt to alter their beliefs to recover market capacity, the correlation in the agent beliefs reduces over time --- due to the different agents' learning updates. Eventually, the system returns to a ``calm" steadier state, with less volatility, supported by the increased diversity of beliefs. 


In summary, the proposed approach captured three key desirable characteristics for a model of economic markets. Specifically, we have shown how the BRATS agents can (i) \textit{converge} to a good average outcome, i.e., equilibrium, on a macro scale, which is punctuated by the (ii) abrupt \textit{emergence of crises} on a micro scale, categorised by spontaneous ``out-of-equilibria deviations" in accordance with actual market dynamics, and in addition, can (iii) recreate \textit{stylised facts} of markets such as clustered volatility.  Importantly, these emergent phenomena occur endogenously, without any external news to the system. The phenomena arise simply due to the interaction among heterogeneous BRATS agents. 

This work adds to the growing literature on the interacting agent hypothesis. There are various approaches which explore this hypothesis using  relatively complex and fine-grained models. For example, heterogeneous agent beliefs can model observed market price dynamics and volatility in the S\&P 500~\cite{lof2015rational}. Likewise, booms and busts in various housing markets can be explained through heterogeneous expectations and beliefs~\cite{bolt2019identifying, zhang2016heterogeneous, bao2019speculators}. 
In contrast, we aimed to capture complex market dynamics within a concise and intuitive model of agent reasoning in a general market setting. In doing so, we verified the hypothesis that the heterogeneity of interacting agents is a crucial factor behind the endogenous emergence of crisis and stylised facts in economic markets.

 Various approaches have been proposed to consider price changes as a result of attendance fluctuations \cite{chen2020complex, kay2004memory, johnson2003financial}. For future work, it may be instructive to incorporate price changes into the analysis presented here, e.g., with the introduction of a market-maker. In addition,  while we have examined entrance into a single market, it would be insightful to explore such findings in a generalised multiple market setting, where agents not only need to decide whether or not to enter, but also which market to enter (if any) in a congestion style game.

In contrast to the long-standing belief that adaptive strategies provides an adequate solution to this class of market games,  we have shown that bounded strategic reasoning can capture salient market dynamics more convincingly. This study highlights that bounded strategic reasoning (``I-think-you-think-they-think...") across heterogeneous agents can explain periods of volatility and abrupt crises emerging in economic markets, even without any external shocks.

\section{Methods}\label{secModel}


\subsection{Background}

 The Quantal Hierarchy (QH) model \cite{evans2021bounded} is based upon a recursive form of the variational free-energy principle, which was proposed as a thermodynamic treatment of bounded rational decision making~\cite{ortega2013thermodynamics}. Agents make a decision $f$ on which action $a \in A$  to take from the available choice set $A$, based upon the utility $U$ of the choice, and the prior beliefs of the agent $p$. Decision making can be represented as state changes, given by the following free energy difference:
\begin{equation}
\label{eqOrtega}
- \Delta F[f[a]] = \sum_{a \in A} f[a] U[a] - \frac{1}{\beta} \sum_{a \in A} f[a] \log \left( \frac{f[a]}{p[a]}\right)
\end{equation} 
which produces the equilibrium distribution for the decision function  $f[a]$:
\begin{equation}\label{eqSingleDecision}
f[a] = \frac{1}{Z} p[a] e^{\beta U[a]}
\end{equation}
where $Z$ is the partition function, 
and the parameter $\beta$ (i.e., inverse temperature) governs the information processing available to an agent. 

The QH model extended this framework to capture recursive  higher-order  reasoning~\cite{evans2021bounded} as pseudo-sequential decision making where agents are limited in the amount of information processing they can perform. 
Specifically, at each level of recursive reasoning $k$, information processing resources are reduced by a discount parameter $\gamma$. Parameter $\gamma$ modulates the beliefs an agent maintains about other agents' decisions. Extending \cref{eqOrtega} to account for (pseudo-)sequential decision making, yields the following recursive free energy difference:
\begin{equation}
- \Delta F[f]  = \sum_{a_{\leq K}} f[a_{\leq K}] \sum_{k=0}^{\infty} \left( U[a_k|a_{<k}] - \frac{1}{\beta \gamma ^{k}} \log  \frac{f[a_k|a_{<k}]}{p[a_k|a_{<k}]} \right)
\end{equation}
where $a_{<k}$ represents the past decisions of the agent. The overall reasoning levels are bound by the recursion depth based on $\gamma$ which ensures, given a simple computational threshold $\epsilon$, that the recursion terminates~\cite{evans2021bounded}. 
The equilibrium solution for the agent decision function $f[a_k|a_{<k}]$ is then given by

\begin{equation}\label{eqDiscountDecision}
  f[a_k \mid a_{<k}] =
\begin{cases}
    \frac{1}{Z_k} \underbrace{p[a_k \mid a_{<k}]}_{\text{Prior Belief}},              & \text{if } \beta \gamma ^{k} < \epsilon
    \\
    \\
    \frac{1}{Z_k} \underbrace{p[a_k \mid a_{<k}]}_{\text{Prior Belief}} \times \underbrace{e ^{\beta U[a_k \mid a_{<k}]}}_{\text{Current Utility}},              & \text{if } \gamma = 0
    \\
    \\
    \frac{1}{Z_k} \underbrace{p[a_k \mid a_{<k}]}_{\text{Prior Belief}} \times \underbrace{Z_{k+1} ^ {1 / \gamma}}_{\text{Future Contribution}} \times \underbrace{e ^{\beta\gamma^{k} U[a_k \mid a_{<k}]}}_{\text{Current Utility}}, & \text{otherwise }
    \\
    \\
\end{cases}
\end{equation}

 The overall processing resources are governed by $\beta\gamma^{k}$, i.e., the information processing abilities $\beta$ are discounted based on  $\gamma$ and the strategic reasoning depth $k$. When $\beta \gamma ^{k} < \epsilon$, the recursion stops since the result  simply echoes the prior belief and no focus is placed on the payoff. In summary, the QH model represents varying levels of higher-order reasoning and captures the expectations about   processing abilities of other agents by varying parameters $\beta$ and $\gamma$. 

\subsection{BRATS approach}
In modelling a market, we assume that each market participant (agent) performs strategic reasoning according to the QH model, i.e.,  recursively but boundedly reasons about  decisions of other agents.   
The reasoning abilities $\beta$ and $\gamma$ are heterogeneously assigned among the participants. All agents start naive, with low information-processing abilities ($\beta \gtrapprox 0$). Over time, agents learn and adjust their beliefs based on observed market outcomes by suitably increasing processing abilities $\beta$ (see below). We refer to this model as the Bounded Rational AdapTive Strategic reasoning (BRATS) approach.

\paragraph{Level-$0$ agent.} The simplest agent type is the level-$0$ agent, or the ``naive player''. These agents have no information-processing abilities ($\beta\gamma^k < \epsilon$), so they make their decisions based on prior beliefs without consideration of other agents' reasoning --- in other words naive players are not strategic agents. These agents simply attend if it would have been profitable to do so previously. 

\paragraph{Learning.} Learning in the BRATS model is represented through increasing reasoning abilities, i.e., increasing $\beta$. This is determined by a learning rate $\eta > 0$, which acts as a linear modifier for $\beta$, as follows: $ \beta_{t+1} = \beta_t + \eta $. All agents start with small $\beta_0$. 
Agents are assigned different rates $\eta$ generated from a given range, representing different learning abilities. This heterogeneous representation provides a simple configuration where varying levels of learning rates can be captured at each timestep across the population of agents.

Parameter $\gamma$ is not directly modified and assumed to be fixed (but different) for all agents. However, increasing $\beta$ implicitly increases the total internal resources used to reason about other agents' reasoning (through $\beta \gamma^k$). 

\subsection{Canonical Model: Adaptive Strategies (AS)}\label{appendixComparison}

The conventional approach to the El Farol bar problem is ``Adaptive Strategies", originally proposed in \cite{arthur1994inductive}, and extended in \cite{fogel1999inductive}. Each market participant $i$ has a set of predictors $\vec{s}_i \in S_i$, creating an ``ecology" of predictors based on the past observations that adapts over time. An individual strategy $\vec{s}_i$ is a vector of weights, where each element in the vector ``determines how the agent believes... the historical data affects the attendance prediction for the current [time]" \cite{wilensky2015introduction}. In contrast to the BRATS approach, the AS model is not a strategic reasoning model. Instead, each agent makes their prediction based on the (public) history, and does not consider the reasoning of other agents at the current timestep. Agents weigh the likelihood of using predictors according to the past success of  this predictor based on a given history, choosing the strategy $\vec{s^*}_i$ which would have done the best in the previous timesteps \cite{wilensky2015introduction}. To implement the AS model for our comparative analysis, we used the model presented in \cite{wilensky2015introduction}, specifically the ``El Farol" model from Chapter 3.

\newpage
\appendix

\section{Experimental Setup}

\subsection{Code Availability}
The complete source code for this paper is available online at: https://github.com/ben-ix/MultiAgentMarkets.

\subsection{Experiment Runs}
All experiments are run for $30$ Monte Carlo simulations to account for stochasticity. The explored range of attendance rates is  $c \in [0.01, 0.02, \dots, 0.99]$. Generally, the marginal cases of $c=0$ and $c=1$ are excluded as trivial, as these correspond to every agent going or staying out. When showing individual plots for $c$'s, typically only $c \in [0.1, 0.2, \dots, 0.9]$ are displayed for readability.

\section{Tail Indices}
The tail-index $\alpha$ produced by the proposed approach falls in the $1-3$ range, as expected by crisis dynamics of emerging ($1 < \alpha < 2$) and established ($2 < \alpha < 4$) markets~\cite{lux2000volatility,lebaron2005extreme}, as shown in \cref{tblTailIndex}. In contrast,  the AS approach generates the $\alpha$'s typically in the range $2.5-6$ (i.e., smaller tails). With noise traders, the $\alpha$'s are also found to be significantly larger, in the range $4.5-7.5$, indicating even smaller tails. These tail indices are visualised in \cref{figViolin}.

\begin{table}[!htb]
\centering
\begin{tabular}{@{}cr|c|ccccccccc@{}}
\toprule
\rowcolor[HTML]{F5F5F5} 
&                &                  & \multicolumn{9}{c}{c} \\
\rowcolor[HTML]{F5F5F5} 

                             &                & $\mu \pm \sigma$ & \textbf{0.1} & \textbf{0.2} & \textbf{0.3} & \textbf{0.4} & \textbf{0.5} & \textbf{0.6} & \textbf{0.7} & \textbf{0.8} & \textbf{0.9}  \\ \midrule
\multirow{3}{*}{{\ul 2.5\%}} & \textbf{BRATS} & 2.3 $\pm$ 0.5    & 1.7 & 1.6 & 2.5 & 2.7 & 2.8 & 2.7 & 2.4 & 1.9 & 2.4 \\
                             & AS             & 4.7 $\pm$ 1.1    & 3.1 & 3.4 & 4.1 & 4.4 & 4.6 & 6.1 & 5.8 & 6.2 & 4.7 \\
                             & Noise          & 7.0 $\pm$ 0.3    & 7.1 & 7.0 & 6.8 & 7.2 & 7.5 & 6.6 & 7.2 & 7.0 & 7.0 \\
                             \midrule
\multirow{3}{*}{{\ul 5\%}}   & \textbf{BRATS} & 1.9 $\pm$ 0.8    & 0.8 & 1.3 & 2.0 & 2.6 & 3.1 & 2.5 & 2.0 & 1.4 & 1.2 \\
                             & AS             & 4.1 $\pm$ 1.0    & 2.7 & 2.9 & 3.5 & 3.8 & 4.2 & 4.9 & 5.1 & 5.4 & 4.6 \\
                             & Noise          & 5.9 $\pm$ 0.2    & 5.9 & 6.2 & 5.7 & 6.1 & 6.2 & 5.8 & 5.9 & 5.8 & 5.9 \\ \midrule
\multirow{3}{*}{{\ul 10\%}}  & \textbf{BRATS} & 1.6 $\pm$ 0.6    & 1.1 & 1.0 & 1.4 & 2.1 & 2.6 & 2.0 & 1.5 & 1.1 & 1.2 \\

                             & AS             & 3.5 $\pm$ 0.8    & 2.3 & 2.6 & 3.0 & 3.4 & 3.6 & 4.2 & 4.1 & 4.5 & 4.1 \\
                             & Noise          & 4.6 $\pm$ 0.1    & 4.6 & 4.9 & 4.6 & 4.8 & 4.6 & 4.5 & 4.6 & 4.6 & 4.6 \\\bottomrule
\end{tabular}
\caption{Average tail-index $\alpha$ computed using the Hill Estimator. Each subtable provides the results for a tail size. Each column for an attendance rate $c$, with the averages across $c$'s presented as the $\mu \pm \sigma$ column. Each cell shows mean estimated tail-index $\alpha$ for a tail size and $c$. A lower $\alpha$ indicates a slower decay, and thus a fatter tail.}\label{tblTailIndex}
\end{table}

\section{Population Heterogeneity}\label{appendixDiversity}
The normalised entropy of the population is used to measure the approximate heterogeneity (diversity) of the agents' beliefs. The resulting entropy is normalised by dividing by the maximal entropy distribution to provide a value between $0$ and $1$, with $0$ indicating no diversity and $1$ indicating maximal diversity.  This formulation allows us to explore the relationship between the population diversity and the resulting attendance volatility, and to explain the potential emergence of periods of high volatility based on population belief changes.

For the BRATS approach (based on QH model), this corresponds to the entropy of agent resources $\beta$'s, with each agent following a behaviour defined by \cref{eqDiscountDecision} at any given time.  Given the set $R$ of all agent resources  $\beta_i$ across $N$ agents ($1 \le i \le N$), the number of bins $|B|$ is chosen according to the Freedman-Diaconis rule for bin width $W$, as follows: 
\begin{equation}
\begin{split}
   W &= \frac{2 \ {\text{IQR}}(R)}{\sqrt[3]{N}} \\
   |B| &= \frac{\max(R)-\min(R)}{W}
   \end{split}
\end{equation}
Then, the entropy is defined across $B$ bins:
\begin{equation}
\begin{split}
    H_B &= -\sum_{b \in B} x_b \log_{2} x_b  \\
    \overline{H}_B &= -\sum_{b \in B} \frac{1}{|B|} \log_{2} \frac{1}{|B|} \\
    H_{\beta} &= H_B / \overline{H}_B \\
\end{split}
\end{equation}
where $x_b$ is the proportion of agent resources $\beta$ distributed within bin $b$. 

For the AS approach, the diversity is quantified by the entropy of the bitstring representation of the chosen strategies $\vec{s^*}_{i}$ across across $N$ agents ($1 \le i \le N$) at a given time. For each agent $i$, in order to convert the strategy $\vec{s^*}_i$ into a bitstring $\vec{v}_i$, we use the following function converting an element $s_{ik} \in \vec{s^*}_i$ to the corresponding element $v_{ik} \in \vec{v}_i$:
\begin{equation}
    v_{ik}(s_{ik}) = 
\begin{cases}
    1,& \text{if } s_{ik}\geq 0\\
    0,              & \text{otherwise}
\end{cases} 
\end{equation}
The bitstring $\vec{v}_i$ is then converted to its decimal representation
\begin{equation}
    D_i = \sum_{j=1}^{|\vec{v}_i|} v_{ij} 2^{|\vec{v}_i|-j}
\end{equation}
where $v_{ij}$ is the $j$'th element in bitstring $\vec{v}_i$ for agent $i$. The decimal representations $D_i$ are then used to compute the entropy. As before, we form the set $D$ of all the decimal representations $D_i$ across $N$ agents and choose the number of bins $|L|$  according to the Freedman-Diaconis rule.  The entropy is computed as follows:
\begin{equation}
\begin{split}
    H_L &= -\sum_{l \in L} x_l  \log_{2} x_l  \\
        \overline{H}_L  &= -\sum_{l \in L} \frac{1}{|L|} \log_{2} \frac{1}{|L|} \\
    H_{S} &= H_L / \overline{H}_L \\
\end{split}
\end{equation}
where $x_l$ is the proportion of decimal representations $D$ distributed within bin $l$.

We note that the utilised measure $H_{\beta}$ quantifies the diversity of \textit{beliefs} of the population, and not the diversity of \textit{information}, as each agent observes the same past historical attendance. For the AS approach, the measure $H_{S}$ quantifies the diversity of the \textit{predictors}. 

\section{Granger Causality}\label{appendixGrangerProcess}
\begin{figure}[!htb]
     \centering
     \begin{subfigure}[b]{0.3\textwidth}
         \centering
         \includegraphics[width=\textwidth]{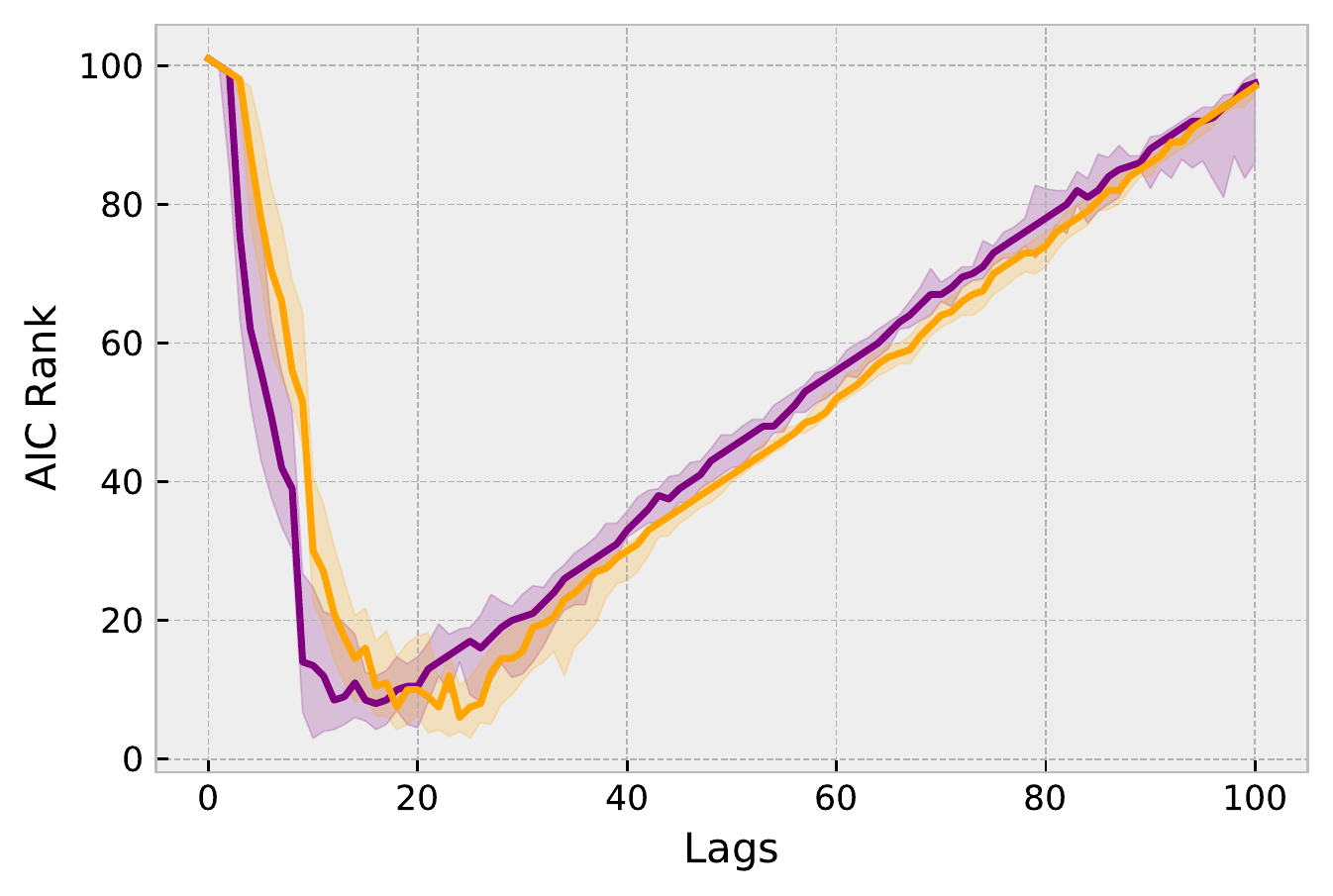}
         \caption{$c=0.1$}
     \end{subfigure}
     \hfill
     \begin{subfigure}[b]{0.3\textwidth}
         \centering
         \includegraphics[width=\textwidth]{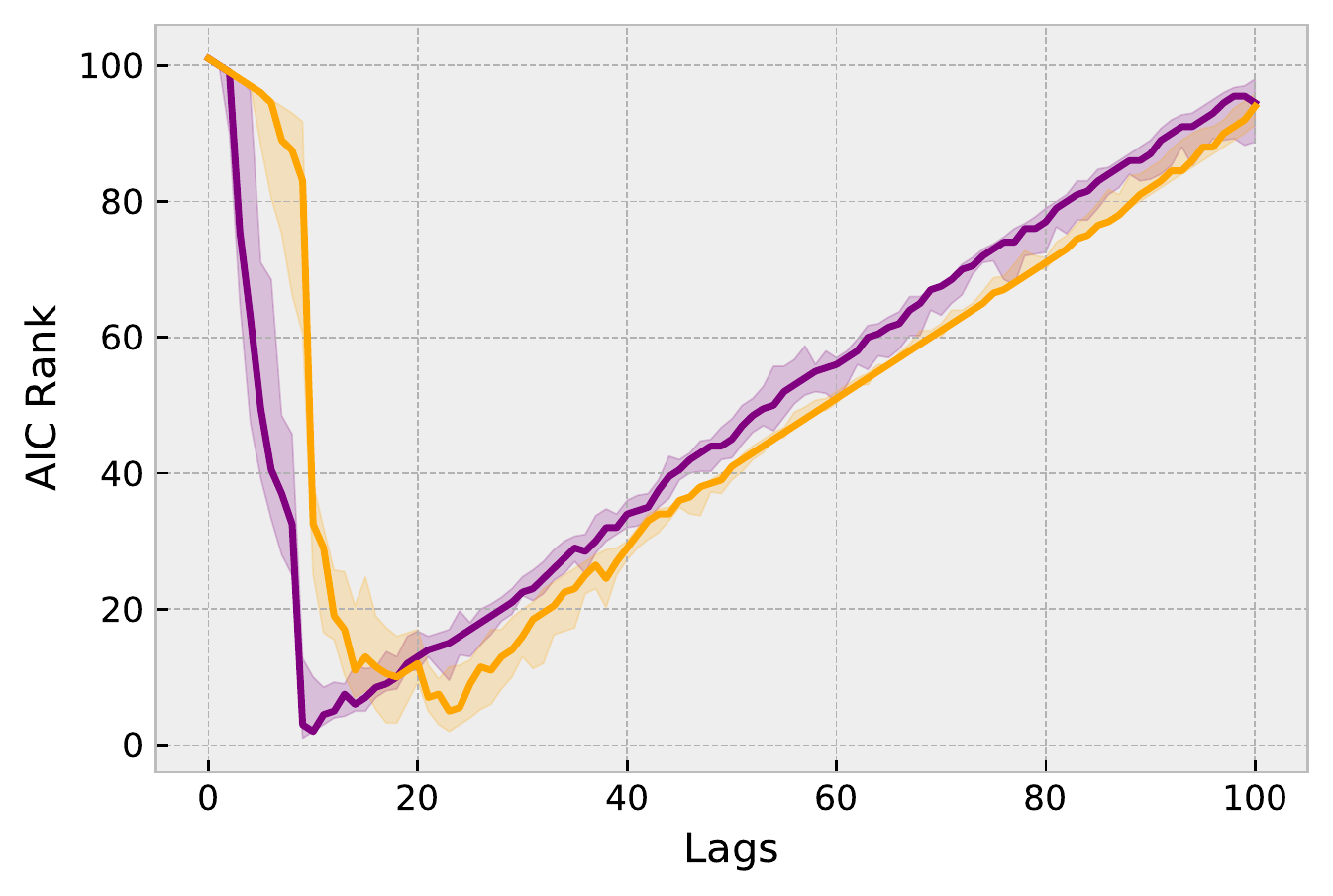}
         \caption{$c=0.2$}
     \end{subfigure}
     \hfill
     \begin{subfigure}[b]{0.3\textwidth}
         \centering
         \includegraphics[width=\textwidth]{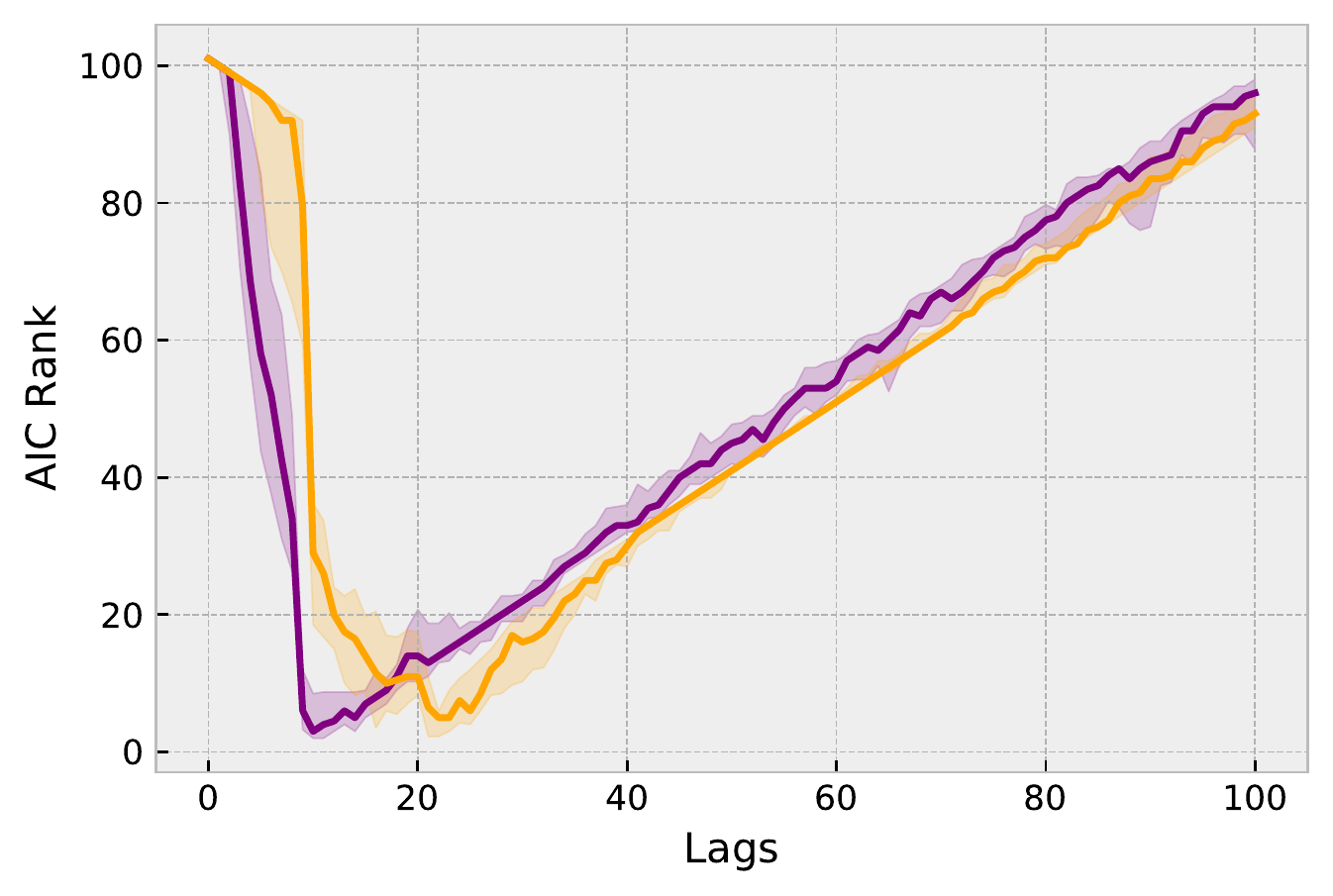}
         \caption{$c=0.3$}
     \end{subfigure}
     \hfill
          \begin{subfigure}[b]{0.3\textwidth}
         \centering
         \includegraphics[width=\textwidth]{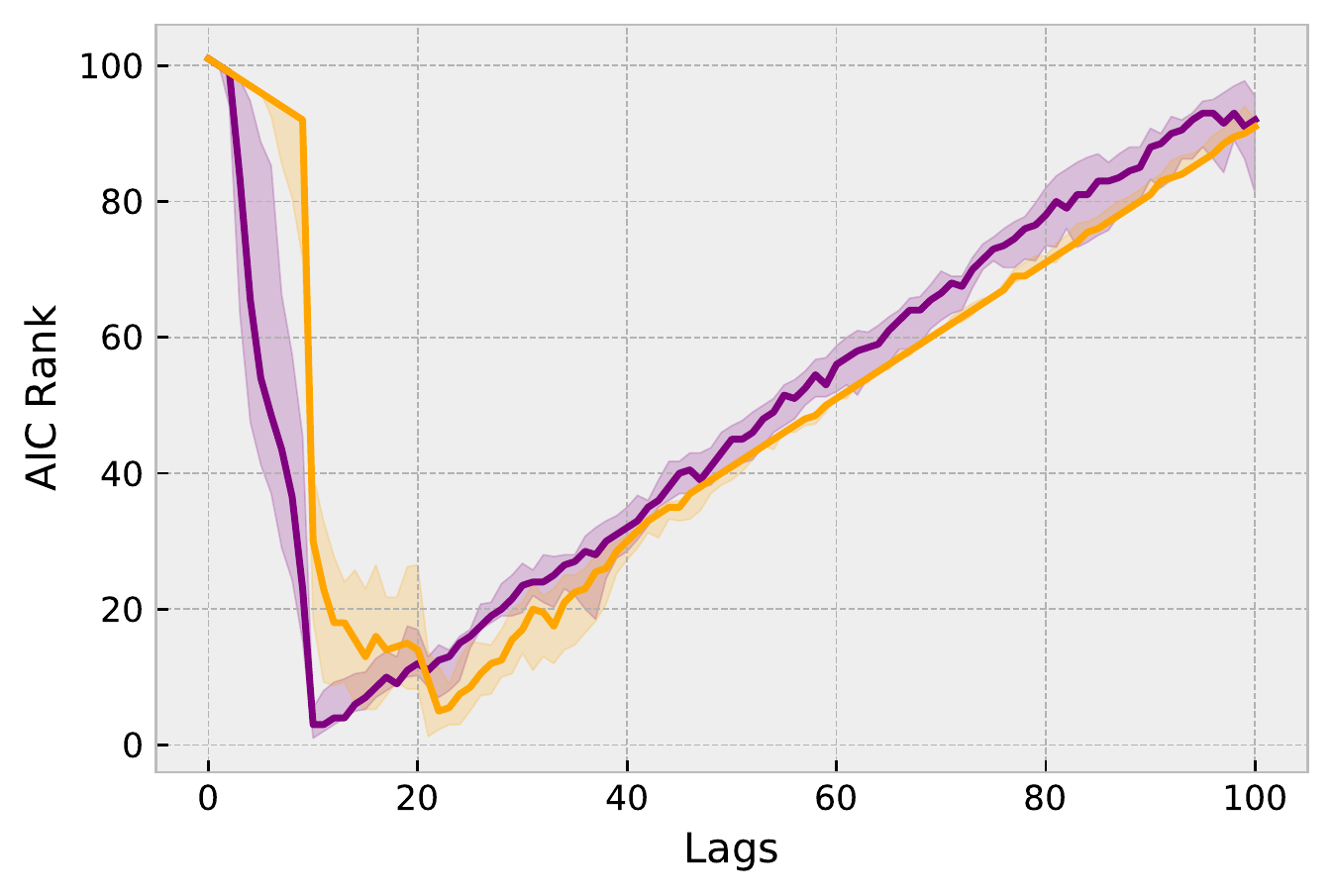}
         \caption{$c=0.4$}
     \end{subfigure}
     \hfill
     \begin{subfigure}[b]{0.3\textwidth}
         \centering
         \includegraphics[width=\textwidth]{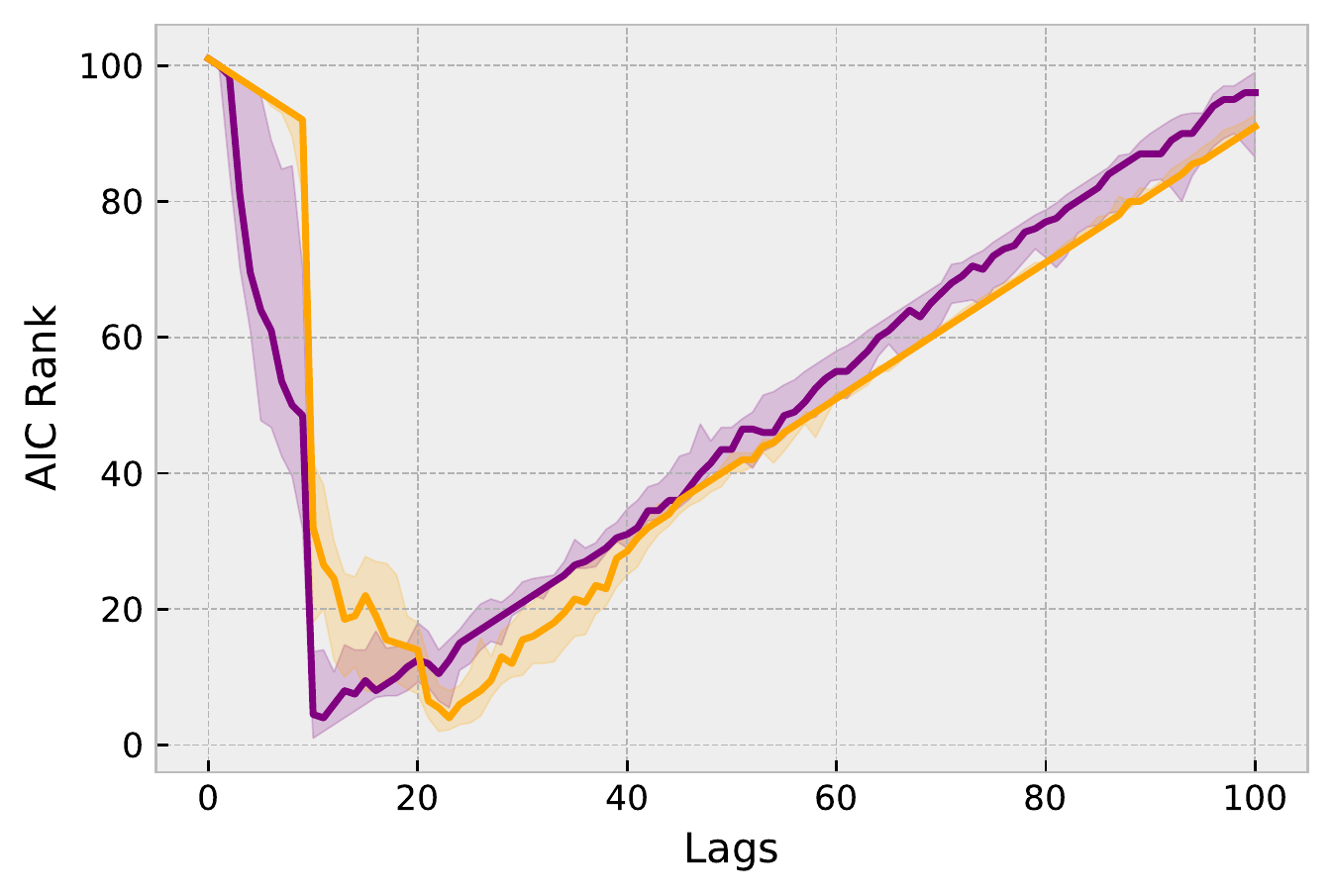}
         \caption{$c=0.5$}
     \end{subfigure}
     \hfill
     \begin{subfigure}[b]{0.3\textwidth}
         \centering
         \includegraphics[width=\textwidth]{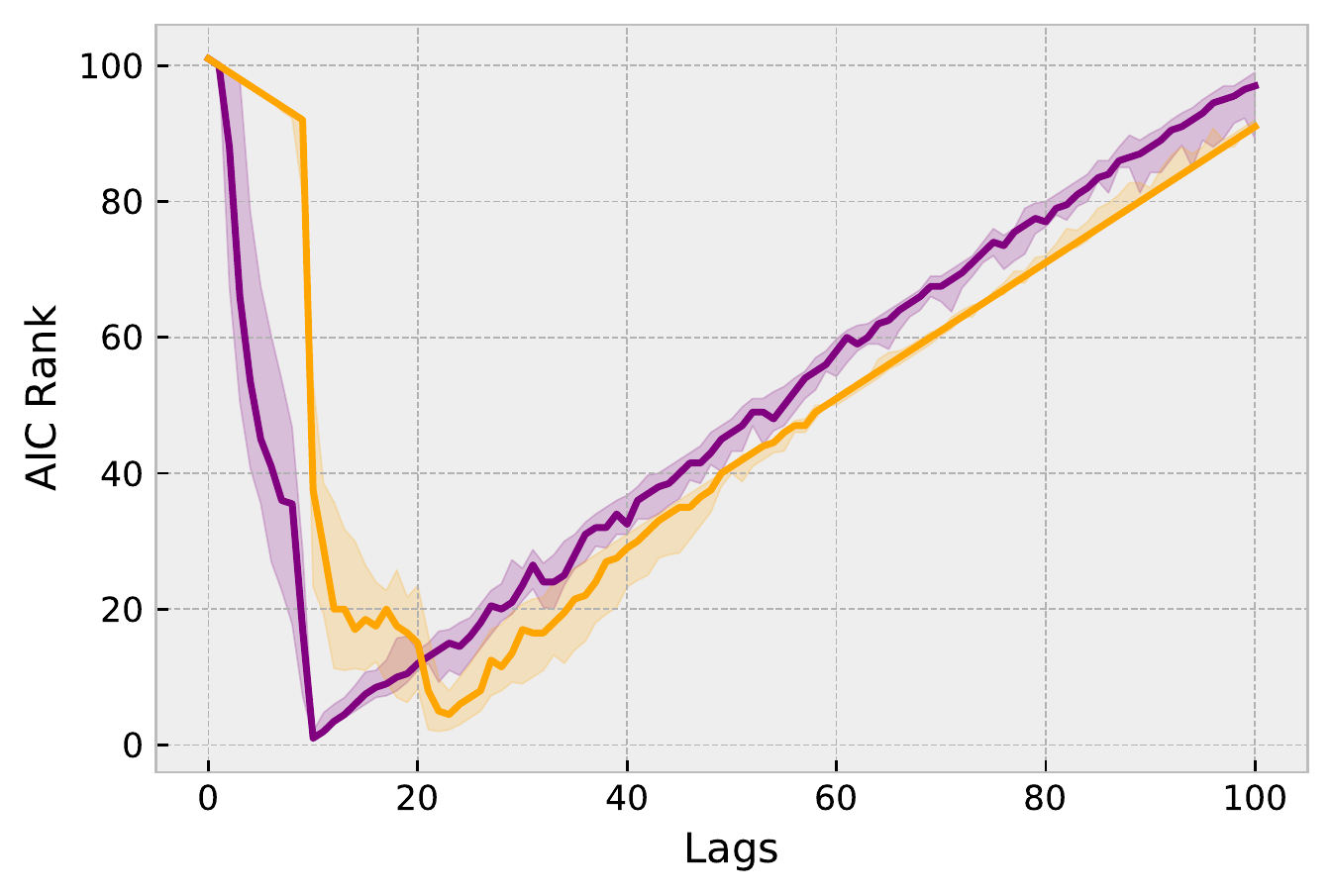}
         \caption{$c=0.6$}
     \end{subfigure}
     \hfill
          \begin{subfigure}[b]{0.3\textwidth}
         \centering
         \includegraphics[width=\textwidth]{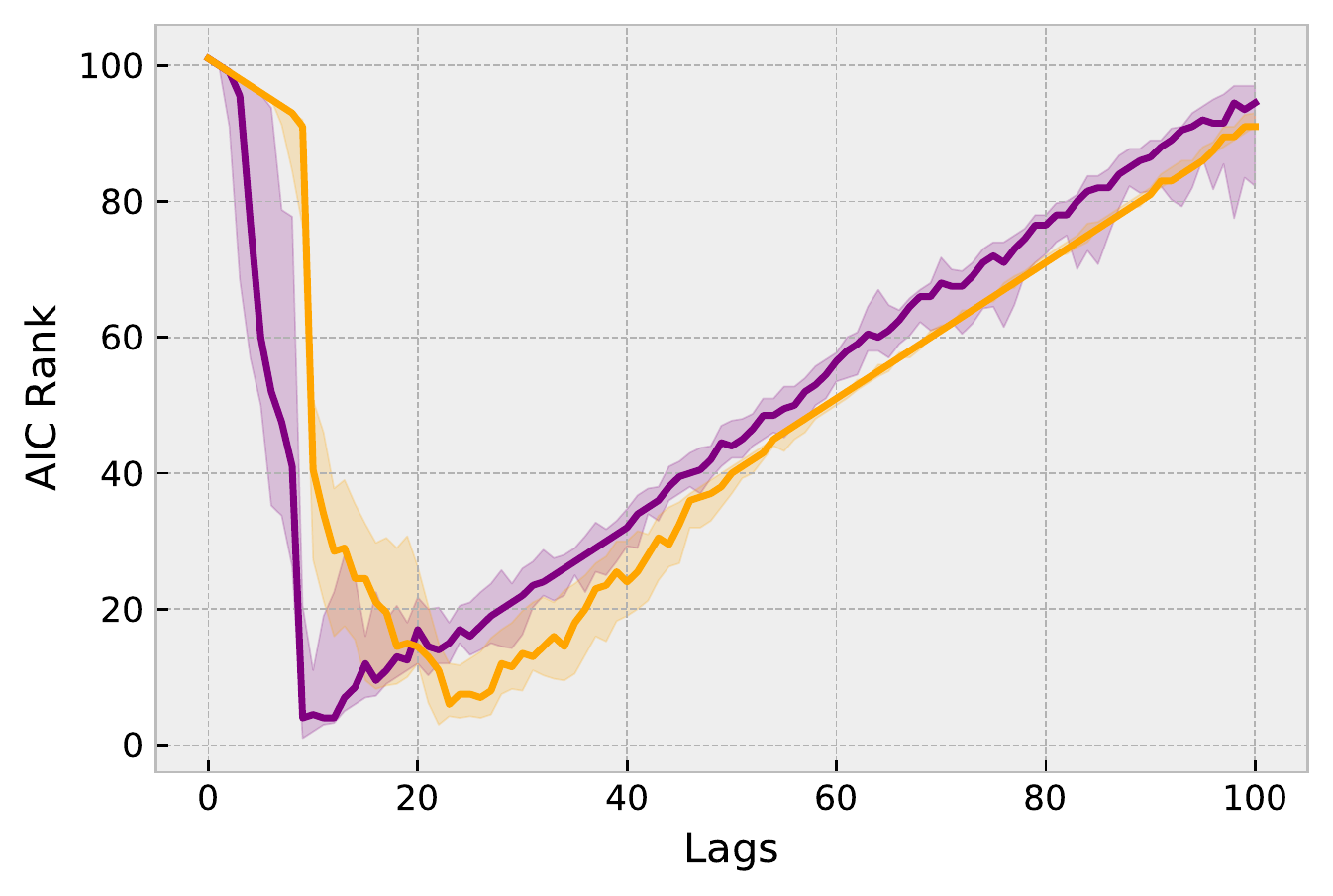}
         \caption{$c=0.7$}
     \end{subfigure}
     \hfill
     \begin{subfigure}[b]{0.3\textwidth}
         \centering
         \includegraphics[width=\textwidth]{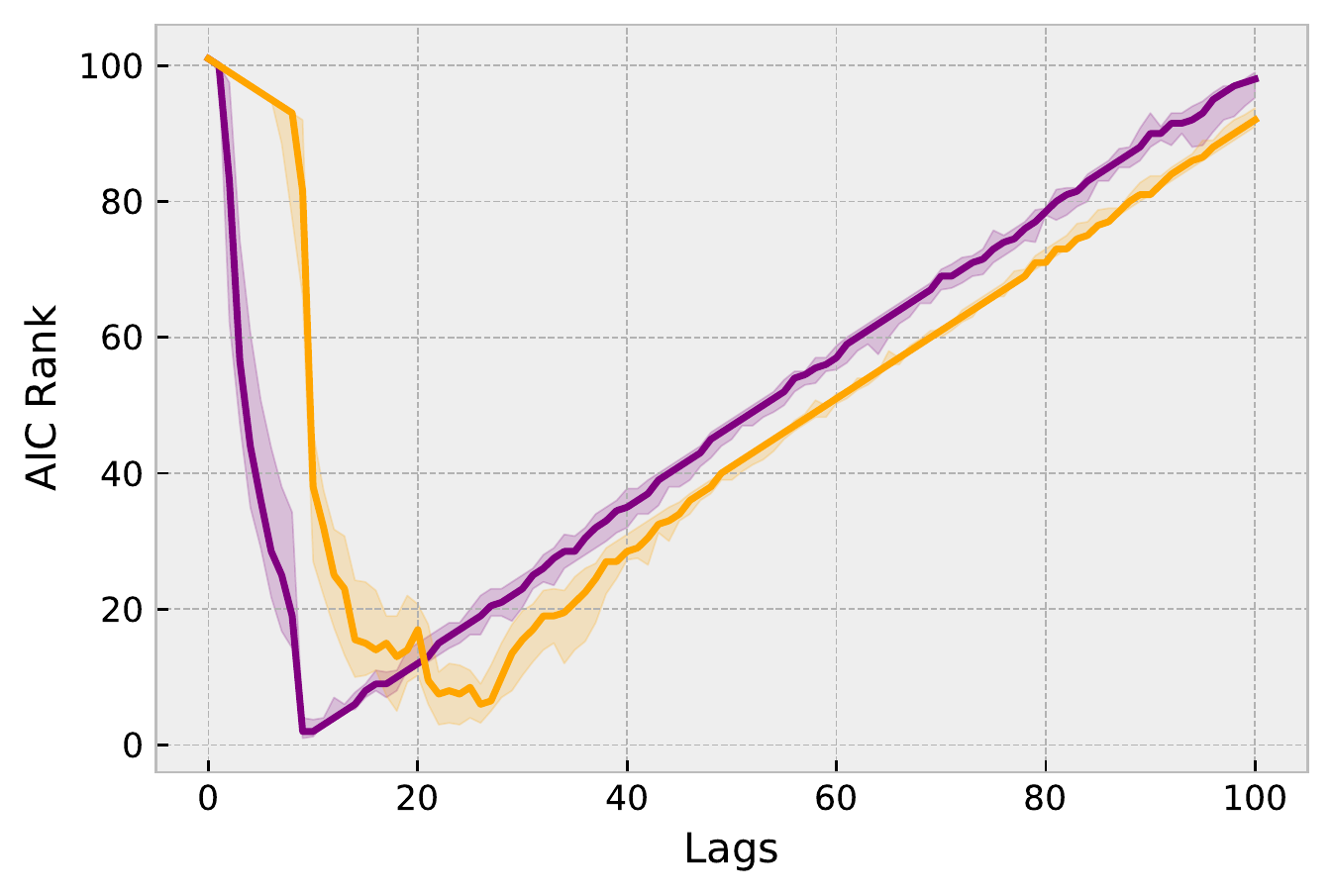}
         \caption{$c=0.8$}
     \end{subfigure}
     \hfill
     \begin{subfigure}[b]{0.3\textwidth}
         \centering
         \includegraphics[width=\textwidth]{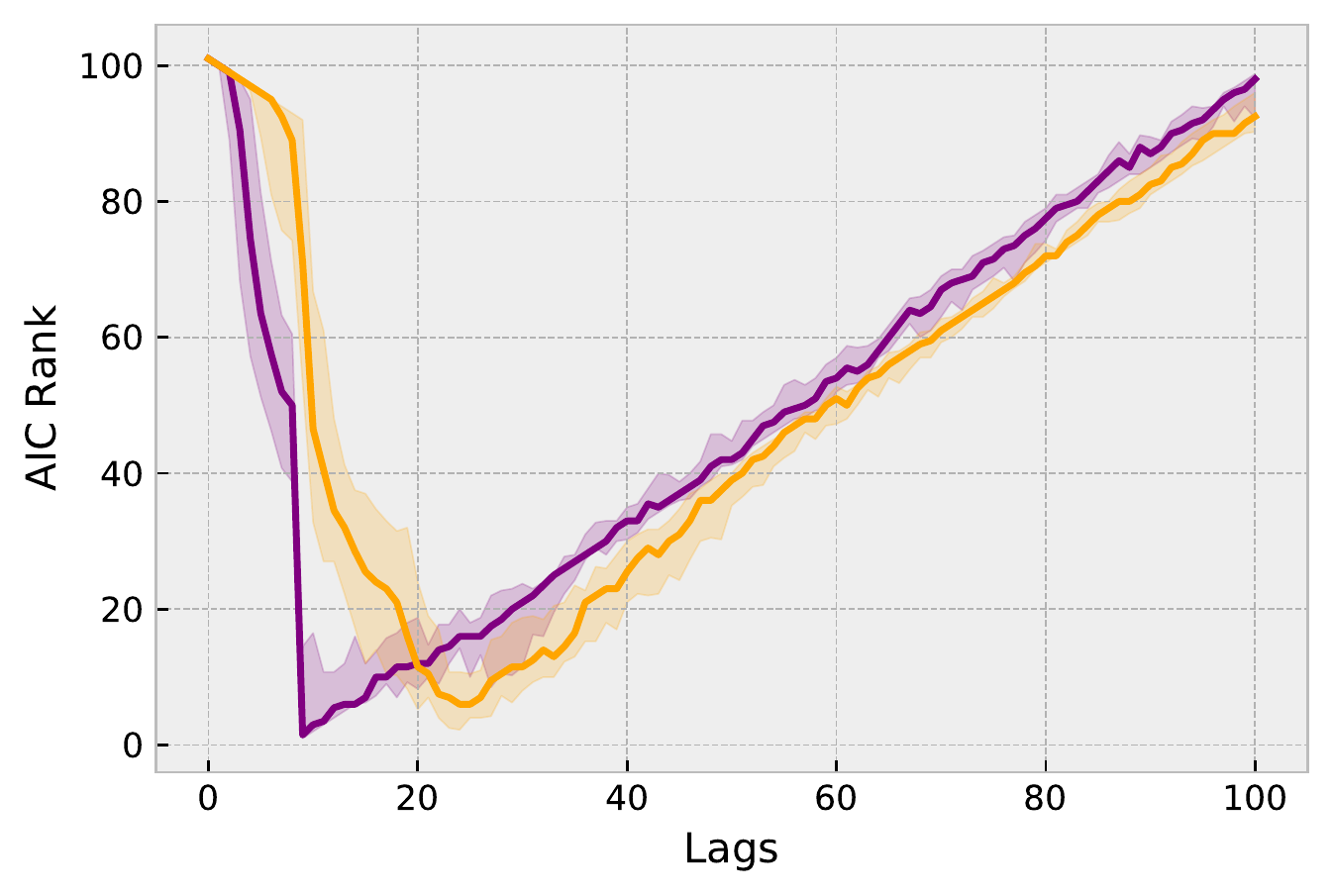}
         \caption{$c=0.9$}
     \end{subfigure}
        \caption{Lag length estimate based on expected rank of AIC, where lowest rank indicates best performance. Purple (orange) represents the proposed BRATS model (adaptive strategies). The darker line indicates the median rank, and the lighter filled region highlights the interquartile range.}
        \label{figLagLength}
\end{figure}

Granger causality is used to test if changes in diversity predict changes in attendance. In other words, this test is used to check if there is information in the (current) diversity about future volatility. To run the Granger causality test, we first consider stationary time series by taking the change in volatility and the change in diversity (first-order differencing). 
The time series are confirmed to have unit-root with the Augmented Dickey-Fuller test, and stationarity with the Kwiatkowski–Phillips–Schmidt–Shin tests. A vector autoregression (VAR) model is then fitted to these time series. The lag-order $L$ of the model is selected by minimising the Akaike information criterion (AIC) \cite{ivanov2005practitioner}. 
This selection process is visualised in \cref{figLagLength}. VAR($L$) is then used to test for Granger causality.  For each $c$, the harmonic mean p-values \cite{wilson2019harmonic} is given to provide an overall significance level across realizations, with a Bonferonni correction to account for multiple tests across $c$.

\begin{table}[!htb]
\centering
\caption{Significance testing for Granger causality between the diversity of agent beliefs and local volatility in attendance in the proposed approach.  All results are significant at the $95\%$ confidence level.}\label{tblGrangerSignficanceProposed}
\begin{tabular}{@{}cccccccccc@{}}
\toprule
\multicolumn{10}{c}{\textbf{Granger Causality (diversity $\to$ volatility)}}                                                                                                                                                                                                                                                                                                                      \\ \midrule
\textbf{c}  & \textbf{0.1}                            & \textbf{0.2}                            & \textbf{0.3}                            & \textbf{0.4}                 & \textbf{0.5}                 & \textbf{0.6}                 & \textbf{0.7}                            & \textbf{0.8}                            & \textbf{0.9}                            \\
\textbf{p*} & \cellcolor[HTML]{78F5A3}\textless{}0.01 & \cellcolor[HTML]{78F5A3}\textless{}0.01 & \cellcolor[HTML]{78F5A3}\textless{}0.01 & \cellcolor[HTML]{78F5A3}0.04 & \cellcolor[HTML]{78F5A3}0.02 & \cellcolor[HTML]{78F5A3}0.02 & \cellcolor[HTML]{78F5A3}\textless{}0.01 & \cellcolor[HTML]{78F5A3}\textless{}0.01 & \cellcolor[HTML]{78F5A3}\textless{}0.01 \\ \bottomrule
\end{tabular}
\end{table}

The null hypothesis $H_0$ for this test is that diversity does not Granger cause volatility. Rejection of $H_0$ confirms if the population heterogeneity can be seen as a leading indicator for volatility in the proposed market game. Failure to reject $H_0$ means that there is no significant predictive information in population diversity about future attendance fluctuations. The resulting p-values from the significance tests are presented in \cref{tblGrangerSignficanceProposed}. The test for Granger causality is significant at the $95\%$ confidence level across all entry rates, indicating we can reject $H_0$ and confirm that diversity change is a leading indicator of volatility in the market game. This result shows that the distribution of agent beliefs is a significant predictor of volatility in the market game.

\newpage
\bibliographystyle{ieeetr}
\bibliography{bib}

\begin{thebibliography}{10}

\bibitem{casson2011origin}
M.~Casson and J.~S. Lee, ``The origin and development of markets: A business
  history perspective,'' {\em Business History Review}, vol.~85, no.~1,
  pp.~9--37, 2011.

\bibitem{sewell2011history}
M.~Sewell, ``History of the efficient market hypothesis,'' research note,
  University College London, January 2011.

\bibitem{fama1970efficient}
E.~F. Fama, ``Efficient capital markets: A review of theory and empirical
  work,'' {\em The journal of Finance}, vol.~25, no.~2, pp.~383--417, 1970.

\bibitem{danielsson2003endogenous}
J.~Danielsson and H.~S. Shin, ``Endogenous risk,'' {\em Modern risk management:
  A history}, pp.~297--316, 2003.

\bibitem{ball2009global}
R.~Ball, ``The global financial crisis and the efficient market hypothesis:
  what have we learned?,'' {\em Journal of Applied Corporate Finance}, vol.~21,
  no.~4, pp.~8--16, 2009.

\bibitem{kirilenko2017flash}
A.~Kirilenko, A.~S. Kyle, M.~Samadi, and T.~Tuzun, ``The flash crash:
  High-frequency trading in an electronic market,'' {\em The Journal of
  Finance}, vol.~72, no.~3, pp.~967--998, 2017.

\bibitem{exuberance}
A.~Smith, T.~Lohrenz, J.~King, P.~R. Montague, and C.~F. Camerer, ``Irrational
  exuberance and neural crash warning signals during endogenous experimental
  market bubbles,'' {\em Proceedings of the National Academy of Sciences},
  vol.~111, no.~29, pp.~10503--10508, 2014.

\bibitem{shiller2000irrational}
R.~C. Shiller, ``Irrational exuberance,'' {\em Philosophy and Public Policy
  Quarterly}, vol.~20, no.~1, pp.~18--23, 2000.

\bibitem{buchanan2009meltdown}
M.~Buchanan, ``Meltdown modelling: could agent-based computer models prevent
  another financial crisis?,'' {\em Nature}, vol.~460, no.~7256, pp.~680--683,
  2009.

\bibitem{farmer2009economy}
J.~D. Farmer and D.~Foley, ``The economy needs agent-based modelling,'' {\em
  Nature}, vol.~460, no.~7256, pp.~685--686, 2009.

\bibitem{hommesStyilized}
C.~H. Hommes, ``Modeling the stylized facts in finance through simple nonlinear
  adaptive systems,'' {\em Proceedings of the National Academy of Sciences},
  vol.~99, no.~suppl\_3, pp.~7221--7228, 2002.

\bibitem{evans2021maximum}
B.~P. Evans and M.~Prokopenko, ``A maximum entropy model of bounded rational
  decision-making with prior beliefs and market feedback,'' {\em Entropy},
  vol.~23, no.~6, p.~669, 2021.

\bibitem{evans2021impact}
B.~P. Evans, K.~Glavatskiy, M.~S. Harr{\'e}, and M.~Prokopenko, ``The impact of
  social influence in {A}ustralian real estate: market forecasting with a
  spatial agent-based model,'' {\em Journal of Economic Interaction and
  Coordination}, pp.~1--53, 2021.

\bibitem{song20222020}
R.~Song, M.~Shu, and W.~Zhu, ``The 2020 global stock market crash: Endogenous
  or exogenous?,'' {\em Physica A: Statistical Mechanics and Its Applications},
  vol.~585, p.~126425, 2022.

\bibitem{wehrli2022classification}
A.~Wehrli and D.~Sornette, ``Classification of flash crashes using the {H}awkes
  (p, q) framework,'' {\em Quantitative Finance}, vol.~22, no.~2, pp.~213--240,
  2022.

\bibitem{mandelbrot1997variation}
B.~B. Mandelbrot, ``The variation of certain speculative prices,'' in {\em
  Fractals and scaling in finance}, pp.~371--418, Springer, 1997.

\bibitem{inoua2020news}
S.~M. Inoua, ``News-driven expectations and volatility clustering,'' {\em
  Journal of Risk and Financial Management}, vol.~13, no.~1, p.~17, 2020.

\bibitem{buchanan2017economists}
M.~Buchanan, ``Economists must broaden their horizons,'' {\em Nature Physics},
  vol.~13, no.~1, pp.~5--5, 2017.

\bibitem{bouchaud2008economics}
J.-P. Bouchaud, ``Economics needs a scientific revolution,'' {\em Nature},
  vol.~455, no.~7217, pp.~1181--1181, 2008.

\bibitem{arthurComplexity}
W.~B. Arthur, ``Complexity and the economy,'' {\em Science}, vol.~284,
  no.~5411, pp.~107--109, 1999.

\bibitem{harre2021complexity}
M.~S. Harr{\'e}, A.~Eremenko, K.~Glavatskiy, M.~Hopmere, L.~Pinheiro,
  S.~Watson, and L.~Crawford, ``Complexity economics in a time of crisis:
  Heterogeneous agents, interconnections, and contagion,'' {\em Systems},
  vol.~9, no.~4, p.~73, 2021.

\bibitem{simon1956rational}
H.~A. Simon, ``Rational choice and the structure of the environment.,'' {\em
  Psychological review}, vol.~63, no.~2, p.~129, 1956.

\bibitem{levitt2008homo}
S.~D. Levitt and J.~A. List, ``Homo economicus evolves,'' {\em Science},
  vol.~319, no.~5865, pp.~909--910, 2008.

\bibitem{camerer2006does}
C.~F. Camerer and E.~Fehr, ``When does `economic man' dominate social
  behavior?,'' {\em Science}, vol.~311, no.~5757, pp.~47--52, 2006.

\bibitem{arthur2021foundations}
W.~B. Arthur, ``Foundations of complexity economics,'' {\em Nature Reviews
  Physics}, vol.~3, no.~2, pp.~136--145, 2021.

\bibitem{kaizoji2000speculative}
T.~Kaizoji, ``Speculative bubbles and crashes in stock markets: an
  interacting-agent model of speculative activity,'' {\em Physica A:
  Statistical Mechanics and its Applications}, vol.~287, no.~3-4, pp.~493--506,
  2000.

\bibitem{hommes2002modeling}
C.~H. Hommes, ``Modeling the stylized facts in finance through simple nonlinear
  adaptive systems,'' {\em Proceedings of the National Academy of Sciences},
  vol.~99, no.~suppl 3, pp.~7221--7228, 2002.

\bibitem{alfarano2005estimation}
S.~Alfarano, T.~Lux, and F.~Wagner, ``Estimation of agent-based models: the
  case of an asymmetric herding model,'' {\em Computational Economics},
  vol.~26, no.~1, pp.~19--49, 2005.

\bibitem{schmitt2017herding}
N.~Schmitt and F.~Westerhoff, ``Herding behaviour and volatility clustering in
  financial markets,'' {\em Quantitative Finance}, vol.~17, no.~8,
  pp.~1187--1203, 2017.

\bibitem{arthur1994inductive}
W.~B. Arthur, ``Inductive reasoning and bounded rationality,'' {\em American
  Economic Review}, vol.~84, no.~2, pp.~406--411, 1994.

\bibitem{casti1996seeing}
J.~L. Casti, ``Seeing the light at {E}l {F}arol: a look at the most important
  problem in complex systems theory,'' {\em Complexity}, vol.~1, no.~5,
  pp.~7--10, 1996.

\bibitem{sellers2020simulating}
M.~W. Sellers, H.~Sayama, and A.~D. Pape, ``Simulating systems thinking under
  bounded rationality,'' {\em Complexity}, vol.~2020, 2020.

\bibitem{st2021network}
S.~St.~Luce and H.~Sayama, ``Network-based phase space analysis of the {E}l
  {F}arol bar problem,'' {\em Artificial Life}, vol.~27, no.~2, pp.~113--130,
  2021.

\bibitem{camerer2004cognitive}
C.~F. Camerer, T.-H. Ho, and J.-K. Chong, ``A cognitive hierarchy model of
  games,'' {\em The Quarterly Journal of Economics}, vol.~119, no.~3,
  pp.~861--898, 2004.

\bibitem{challet2001stylized}
D.~Challet, M.~Marsili, and Y.-C. Zhang, ``Stylized facts of financial markets
  and market crashes in minority games,'' {\em Physica A: Statistical Mechanics
  and its Applications}, vol.~294, no.~3-4, pp.~514--524, 2001.

\bibitem{challet2004shedding}
D.~Challet, M.~Marsili, and G.~Ottino, ``Shedding light on {E}l {F}arol,'' {\em
  Physica A: Statistical Mechanics and Its Applications}, vol.~332,
  pp.~469--482, 2004.

\bibitem{evans2021bounded}
B.~P. Evans and M.~Prokopenko, ``Bounded rationality for relaxing best response
  and mutual consistency: The quantal hierarchy model of decision-making,''
  {\em arXiv preprint arXiv:2106.15844}, 2021.

\bibitem{shubik2011farol}
M.~Shubik, ``{E}l {F}arol revisited: A note on emergence, game theory, and
  society,'' {\em Complexity}, vol.~16, no.~6, pp.~62--65, 2011.

\bibitem{whitehead2008farol}
D.~Whitehead {\em et~al.}, ``The {E}l {F}arol bar problem revisited:
  Reinforcement learning in a potential game,'' {\em ESE discussion papers},
  vol.~186, 2008.

\bibitem{kay2004memory}
R.~Kay and N.~F. Johnson, ``Memory and self-induced shocks in an evolutionary
  population competing for limited resources,'' {\em Physical Review E},
  vol.~70, no.~5, p.~056101, 2004.

\bibitem{guzman2020towards}
M.~Guzman and J.~E. Stiglitz, ``Towards a dynamic disequilibrium theory with
  randomness,'' {\em Oxford Review of Economic Policy}, vol.~36, no.~3,
  pp.~621--674, 2020.

\bibitem{jefferies2003anatomy}
P.~Jefferies, D.~Lamper, and N.~F. Johnson, ``Anatomy of extreme events in a
  complex adaptive system,'' {\em Physica A: Statistical Mechanics and its
  Applications}, vol.~318, no.~3-4, pp.~592--600, 2003.

\bibitem{johnson2003financial}
N.~F. Johnson, P.~Jefferies, P.~M. Hui, {\em et~al.}, ``Financial market
  complexity,'' {\em OUP Catalogue}, 2003.

\bibitem{hill1975simple}
B.~M. Hill, ``A simple general approach to inference about the tail of a
  distribution,'' {\em The annals of statistics}, pp.~1163--1174, 1975.

\bibitem{lebaron2005extreme}
R.~Samanta, B.~LeBaron, {\em et~al.}, ``Extreme value theory and fat tails in
  equity markets,'' tech. rep., Society for Computational Economics, 2005.

\bibitem{lux2000volatility}
T.~Lux and M.~Marchesi, ``Volatility clustering in financial markets: a
  microsimulation of interacting agents,'' {\em International journal of
  theoretical and applied finance}, vol.~3, no.~04, pp.~675--702, 2000.

\bibitem{brunnermeier2016bubbles}
M.~K. Brunnermeier, ``Bubbles,'' in {\em Banking Crises}, pp.~28--36, Springer,
  2016.

\bibitem{garber1989tulipmania}
P.~M. Garber, ``Tulipmania,'' {\em Journal of political Economy}, vol.~97,
  no.~3, pp.~535--560, 1989.

\bibitem{mackay2012extraordinary}
C.~Mackay, {\em Extraordinary popular delusions and the madness of crowds}.
\newblock Simon and Schuster, 2012.

\bibitem{keynes1937general}
J.~M. Keynes, ``The general theory of employment,'' {\em The quarterly journal
  of economics}, vol.~51, no.~2, pp.~209--223, 1937.

\bibitem{koppl2002all}
R.~Koppl and J.~Barkley Rosser~Jr, ``All that {I} have to say has already
  crossed your mind,'' {\em Metroeconomica}, vol.~53, no.~4, pp.~339--360,
  2002.

\bibitem{fogel1999inductive}
D.~B. Fogel, K.~Chellapilla, and P.~J. Angeline, ``Inductive reasoning and
  bounded rationality reconsidered,'' {\em IEEE transactions on evolutionary
  computation}, vol.~3, no.~2, pp.~142--146, 1999.

\bibitem{bell2003coordination}
A.~M. Bell, W.~A. Sethares, and J.~A. Bucklew, ``Coordination failure as a
  source of congestion in information networks,'' {\em IEEE Transactions on
  Signal Processing}, vol.~51, no.~3, pp.~875--885, 2003.

\bibitem{kurz2001endogenous}
M.~Kurz and M.~Motolese, ``Endogenous uncertainty and market volatility,'' {\em
  Economic Theory}, vol.~17, no.~3, pp.~497--544, 2001.

\bibitem{rand2007farol}
W.~Rand and F.~Stonedahl, ``The {E}l {F}arol bar problem and computational
  effort: Why people fail to use bars efficiently,'' {\em Northwestern
  University, Evanston, IL}, 2007.

\bibitem{lof2015rational}
M.~Lof, ``Rational speculators, contrarians, and excess volatility,'' {\em
  Management Science}, vol.~61, no.~8, pp.~1889--1901, 2015.

\bibitem{bolt2019identifying}
W.~Bolt, M.~Demertzis, C.~Diks, C.~Hommes, and M.~Van Der~Leij, ``Identifying
  booms and busts in house prices under heterogeneous expectations,'' {\em
  Journal of Economic Dynamics and Control}, vol.~103, pp.~234--259, 2019.

\bibitem{zhang2016heterogeneous}
H.~Zhang, Y.~Huang, and H.~Yao, ``Heterogeneous expectation, beliefs evolution
  and house price volatility,'' {\em Economic Modelling}, vol.~53,
  pp.~409--418, 2016.

\bibitem{bao2019speculators}
T.~Bao and C.~Hommes, ``When speculators meet suppliers: Positive versus
  negative feedback in experimental housing markets,'' {\em Journal of Economic
  Dynamics and Control}, vol.~107, p.~103730, 2019.

\bibitem{chen2020complex}
L.~Chen, ``Complex network minority game model for the financial market
  modeling and simulation,'' {\em Complexity}, vol.~2020, 2020.

\bibitem{ortega2013thermodynamics}
P.~A. Ortega and D.~A. Braun, ``Thermodynamics as a theory of decision-making
  with information-processing costs,'' {\em Proceedings of the Royal Society A:
  Mathematical, Physical and Engineering Sciences}, vol.~469, no.~2153,
  p.~20120683, 2013.

\bibitem{wilensky2015introduction}
U.~Wilensky and W.~Rand, {\em An introduction to agent-based modeling: modeling
  natural, social, and engineered complex systems with NetLogo}.
\newblock Mit Press, 2015.

\bibitem{ivanov2005practitioner}
V.~Ivanov and L.~Kilian, ``A practitioner's guide to lag order selection for
  var impulse response analysis,'' {\em Studies in Nonlinear Dynamics \&
  Econometrics}, vol.~9, no.~1, 2005.

\bibitem{wilson2019harmonic}
D.~J. Wilson, ``The harmonic mean p-value for combining dependent tests,'' {\em
  Proceedings of the National Academy of Sciences}, vol.~116, no.~4,
  pp.~1195--1200, 2019.

\end{thebibliography}

\end{document}